\documentclass[reprint,twocolumn,superscriptaddress,nofootinbib,aps,pre,floatfix]{revtex4-1}
\pdfoutput=1
\usepackage{graphicx}
\usepackage[english]{babel}
\usepackage{amsmath}
\usepackage{amssymb}
\usepackage[usenames,dvipsnames]{color}
\usepackage{url}
\usepackage{bbm}
\usepackage{multirow}
\usepackage{hyperref}
\usepackage{chngcntr}

\newcommand{\norm}[1]{\|#1 \|}
\newcommand{\CE}{\textit{C.~elegans}}
\newcommand{\VI}{\textit{VI}} 
\newcommand{\VIt}{$\langle \textit{VI}(t) \rangle \,$}        
\newcommand{\VItt}{$\textit{VI}(t,t') \,$}

\newcommand{\CV}{\textit{CV}}

\newcommand{\abs}[1]{\left|#1\right|}

\renewcommand{\vec}[1]{\mathbf{#1}}

\begin{document}

\title{Flow-based network analysis  of the \textit{Caenorhabditis elegans} connectome}
\author{Karol A. Bacik} \email[]{karol.bacik13@imperial.ac.uk}
\affiliation{Department of Mathematics, Imperial College London, 
    London SW7 2AZ, United Kingdom}
\author{Michael T. Schaub} \email[]{michael.schaub@unamur.be}
\affiliation{Department of Mathematics, Imperial College London, 
    London SW7 2AZ, United Kingdom}
\affiliation{naXys \& Department of Mathematics, University of Namur, 
    B-5000 Namur, Belgium}
\affiliation{ICTEAM, Universit\'e catholique de Louvain, 
    B-1348 Louvain-la-Neuve, Belgium}    
\author{Mariano Beguerisse-D\'\i az} \email[]{beguerisse@maths.ox.ac.uk}
\affiliation{Department of Mathematics, Imperial College London, 
    London SW7 2AZ, United Kingdom}
\author{Yazan N. Billeh}
\affiliation{Computation and Neural Systems Program, California Institute of Technology, CA 91125 Pasadena, USA}   
\author{Mauricio Barahona} 
\email[]{m.barahona@imperial.ac.uk}
\affiliation{Department of Mathematics, Imperial College London,  
    London SW7 2AZ, United Kingdom}
\date{\today}

\begin{abstract}
    We exploit flow propagation on the directed neuronal network of the nematode~\CE~ to reveal dynamically relevant features of its connectome.  We find flow-based groupings of neurons at different levels of granularity, which we relate to functional and anatomical constituents of its nervous system.  A systematic \textit{in silico} evaluation of the full set of single and double neuron ablations is used to identify deletions that induce the most severe disruptions of the multi-resolution flow structure.  Such ablations are linked to functionally relevant neurons, and suggest potential candidates for further \textit{in vivo} investigation.  In addition, we use the directional patterns of incoming and outgoing network flows at all scales to identify flow profiles for the neurons in the connectome, without pre-imposing \textit{a priori} categories. The four flow roles identified are linked to signal propagation motivated by biological input-response scenarios.
\end{abstract}

\maketitle

\section*{Author Summary}
One of the goals of systems neuroscience is to elucidate the
relationship between the structure of neuronal networks and the
functional dynamics that they implement.  An ideal model organism to
study such interactions is the roundworm \CE, which not only has a
fully mapped connectome, but has also been the object of extensive
behavioural, genetic and neurophysiological experiments.  Here we
present an analysis of the neuronal network of~\CE~from a dynamical
flow perspective.  Our analysis reveals a multi-scale organisation of
the signal flow in the network linked to anatomical and functional
features of neurons and identifies different neuronal roles
in relation to signal propagation.  We use our computational
framework to explore biological input-response scenarios as well as
exhaustive \textit{in silico} ablations, which we relate to
experimental findings reported in the literature.

\section{Introduction}
The nematode \textit{Caenorhabditis elegans} has been used as a model
organism in the life sciences for half a century~\cite{celegans2}, and
considerable effort has been devoted to elucidate the properties of
its nervous system in relation to functional behaviour.  The
\CE~connectome was originally charted in 1986 by White \textit{et
  al}~\cite{white1986} and has been further refined by analysis and
experiments~\cite{Hall1991}, most recently in the work of Varshney
\textit{et al}~\cite{varshney2011}.  Using experimental techniques
such as laser ablations, calcium imaging, optogenetics and
sonogenetics, researchers have examined functional properties of
individual neurons in connection with motion, learning, or information
processing and
integration~\cite{chalfie1985,wakabayashi2004,li2011,Nagel2005,Ibsen2015}.
Other studies have quantified the characteristics of the motion of
\CE, and how these change upon genetic
mutations~\cite{Stephens2008,Yemini2013,Brown2013}.

With the increased availability of data from such experiments, there
is a need to integrate current knowledge about individual neurons into
a comprehensive picture of how the network of neurons
operates~\cite{white1986,chen2006,varshney2011}.  A number of studies
have reported network characteristics of the \CE~connectome: it is a
small-world network~\cite{Watts1998} satisfying mathematical criteria
of efficiency~\cite{Kim2014}, with a heavy-tailed degree
distribution~\cite{Barabasi1999} and a core-set of highly-connected,
`rich club' neurons~\cite{towlson2013}.  Furthermore, the analysis of
modules in the network has shown that certain strongly coupled
clusters of neurons can be linked to biological
functions~\cite{majewska2001,arenas2008,pan2010,sohn2011,pavlovic2014}.
Such observations suggest that a system-wide analysis of the
connectome can provide valuable functional information.  However,
finding simplified mesoscale descriptions that coherently aggregate
how information propagates in the directed connectome across multiple
scales remains a challenge~\cite{Sporns2015}.

In this work, rather than focusing on structural features of the
network, we analyse the directed and weighted \CE~connectome from a
dynamics-based (more specifically, flow-based) perspective.
Using dynamics to probe the relationship between the structure and
function of a system has become a valuable tool in many
settings~\cite{Lambiotte2014,Jeub2015,Sporns2015}.  In particular,
dynamics-based approaches have been successfully used to study brain
networks (e.g., fMRI and DSI data~\cite{Betzel2013,Misic2015,Lizier2011}).
For an in depth discussion of network-theoretic methods 
in neuroscience see the extensive reviews~\cite{Bullmore2009,Sporns2011,Sporns2015}.
For an overview on dynamical methods for network analysis we refer the reader to Refs.~\cite{Delvenne2013,Lambiotte2014,Jeub2015} and the literature cited therein.

 Our methods use diffusive processes on graphs
as a simple means to link features of the directed network and
propagation dynamics.  While diffusive flow is a simplification of the
actual propagation in the nervous system of \CE, we can still gain
insight into network properties of dynamical
interest~\cite{varshney2011}.  We exploit these ideas in two ways.
Firstly, we investigate flow-based partitions of the connectome across
multiple scales using the Markov Stability (MS) framework for
community detection~\cite{Delvenne2010,Delvenne2013, Lambiotte2014}.
Our analysis detects subgroups of neurons that retain diffusive flows
over particular time scales~\cite{Schaub2012} taking into account edge
directionality~\cite{Lambiotte2014, Beguerisse2014}.  We then mimic
neuronal ablations computationally, and check \emph{all} possible
single and double ablations in the connectome to detect those that are most
disruptive of the flow organisation.  Secondly, we extract alternative
information of the directed network flows through the Role Based
Similarity (RBS) framework~\cite{Cooper2010,
  Cooper2010a,Beguerisse2013}. Without pre-imposing categories
\textit{a priori}, RBS classifies neurons into flow roles, i.e., classes of
neurons with similar asymmetric patterns of incoming and outgoing
network flows at all scales, which are directly
extracted from the network.  Finally, we mimic `stimulus-response'
experiments~\cite{chalfie1985,li2011,hillard2002}, in which signals
propagate through the network starting from well-defined sets of input
neurons linked to particular biological stimuli.  The ensuing time courses of
neuronal flows reveal features of information processing in \CE, in
relation to the obtained flow roles. Our computational analyses are
consistent with experimental findings, suggesting that our framework
can provide guidance towards the identification of potential neuronal
targets for further {\it in vivo} experiments.

\section{Results}
\begin{figure*}[tb!] 
    \includegraphics{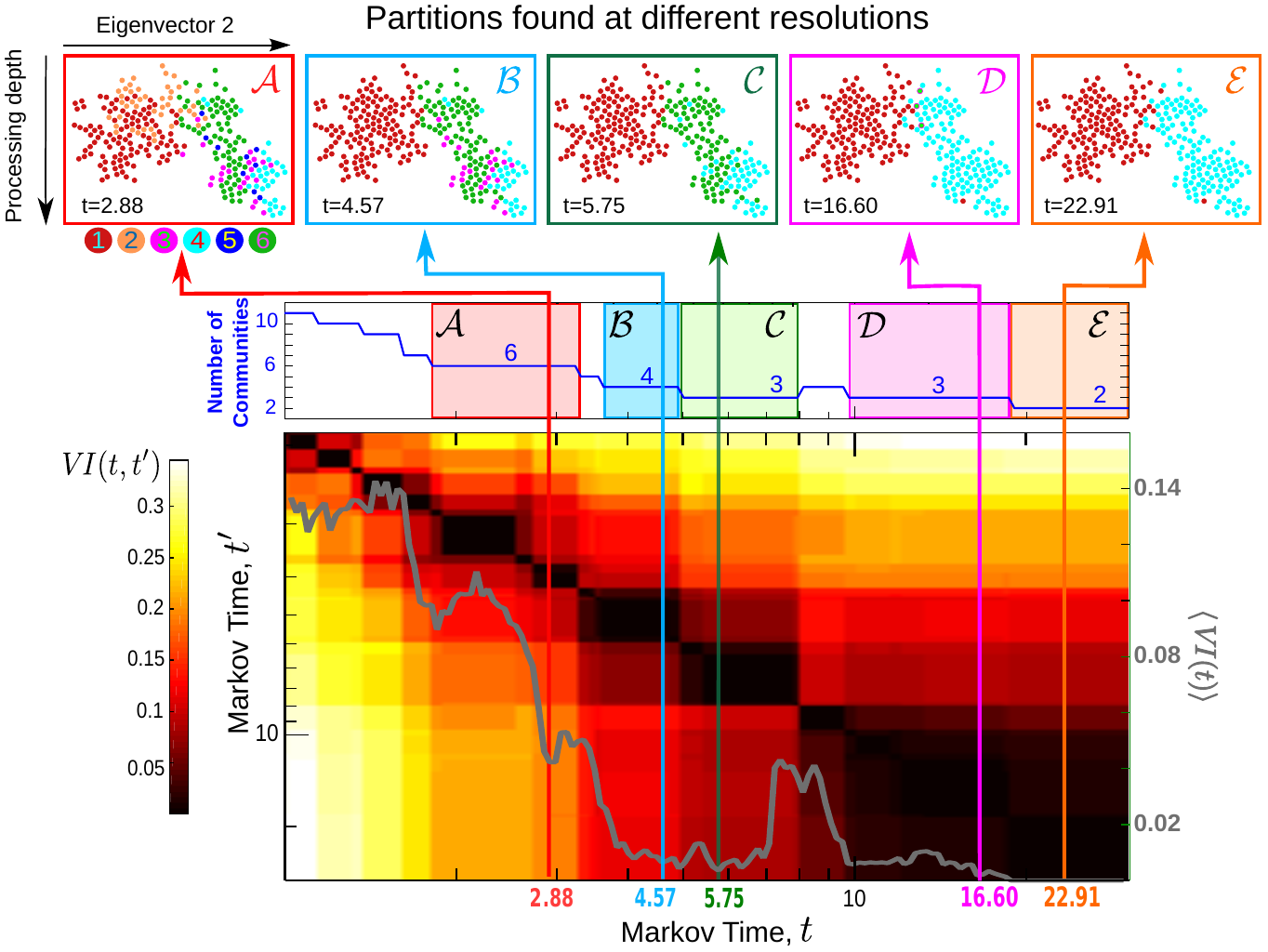} 
    \caption{\textbf{Flow-based multiscale partitioning of the
        connectome of \CE.}  Using Markov Stability, we detect
      flow-based partitions in this directed network at all scales.
      Here we show the medium to coarse Partitions $\mathcal A$ to
      $\mathcal E$ (top panel), found as optimal at the indicated
      Markov time intervals (see Fig.~\ref{S1_Fig} for the full sweep of
      Markov times).  The Markov Time intervals corresponding 
      to different robust partitions are indicated by different colour boxes.      
      Partitions $\mathcal A$-$\mathcal E$      
      are persistent, as signalled by their robustness over extended time plateaux 
      in \VItt (heatmap in bottom panel), 
      and robust with respect to the optimisation, as
      signalled by dips in the variation of information \VIt (grey line in bottom panel).  }
    \label{fig:Stability} 
\end{figure*}

Our analysis uses the \CE~data published in Ref.~\cite{varshney2011}
(see \url{www.wormatlas.org/neuronalwiring.html}). To represent the
\CE~connectome, we use the two-dimensional network layout given
by~\cite{varshney2011}, i.e., neurons are placed on the plane
according to their normalised Laplacian eigenvector ($x$-axis)
and processing depth ($y$-axis), as seen in Fig.~\ref{fig:Stability} (top
panel).  We study the largest weakly-connected component of this
network, which contains 279 neurons with 6394 chemical synapses
(directed) and 887 gap junctions (bidirectional).
Reference~\cite{varshney2011} also provides the position of the soma
of each neuron along the body of the worm, and classifies each neuron
as either sensory (S), interneuron (I) or motor~(M).

\subsection{Flow-based partitioning reveals multi-scale organisation
  of the connectome}
\label{sec:stabilityres}

To reveal the multi-scale flow organisation of the \CE~connectome, we
use the Markov Stability (MS) framework described in
Sec.~`\ref{sec:stability}'.  Conceptually, MS can be understood as
follows.  Imagine that a drop of ink (signal) is placed on a node and
begins to diffuse along the edges of the graph.  If the graph lacks
structural organisation (e.g., random), the ink diffuses isotropically
and rapidly reaches its stationary distribution.  However, the graph
might contain subgraphs in which the flow is trapped for longer than
expected, before diffusing out towards stationarity.  These groups of
nodes constitute dynamical, flow-retaining communities in the graph,
usually signifying a strong dynamic coherence within the group
and a weaker coherence with the rest of the network.  If we allow the ink to diffuse
just for a short time, then only small communities are detected, for
the diffusion cannot explore the whole extent of the network.  If we
observe the process for a longer time, the ink reaches larger parts of
the network, and the flow communities thus become larger.  By employing
dynamics, and in particular by scanning across time, MS can thus
detect cohesive node groupings at different levels of
granularity~\cite{Schaub2012, Lambiotte2014, Billeh2014}.  In this
sense, the time of the diffusion process, denoted \textit{Markov time}
in the following, acts as a resolution parameter.

\begin{figure*}[tbh!] 
    \centering
    \includegraphics{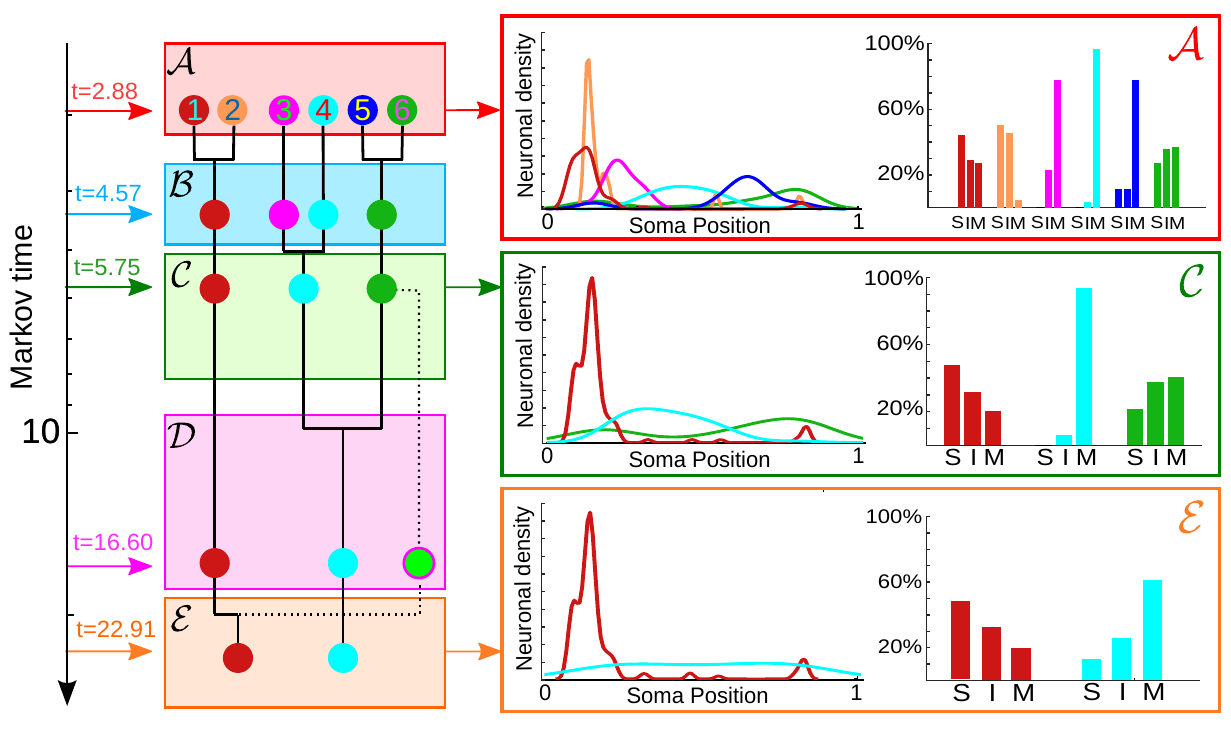} 
    \caption{\textbf{Community structure and biological features.}
      Left: As indicated by the dendrogram, the partitions obtained
      have a quasi-hierarchical organisation. The dotted line
      indicates that the light green community at $t=16.60$ does 
      not result from a hierarchical merging.  Middle: The smoothed
      spatial densities of neurons in each community for the different
      partitions show how the communities are spatially grouped
      according to soma positions along a longitudinal axis
      normalised between 0 and 1. The merging of
      groups over Markov time largely retains this spatial structure.
      Right: The percentages of sensory (S), inter- (I) and motor (M)
      neurons in each community show functional segregation in the
      groupings.  }
    \label{fig:Stability_soma} 
\end{figure*}

The flow-based community structure of the \CE~connectome at medium to
coarse levels of resolution is shown in Fig.~\ref{fig:Stability}.  The
full scan across all Markov times is shown in Fig.~\ref{S1_Fig} and
section~\ref{S1_Data}.  As described above, the partitions become coarser as the
Markov time $t$ increases, from the finest possible partition, in
which each node forms its own community, to the dominant bi-partition
at long Markov times.  The sequence of partitions exhibits an {\it
  almost hierarchical} structure, with a strong spatial localisation
linked to functional and organisational circuits
(see~Fig.~\ref{fig:Stability_soma} and Fig.~\ref{S2_Fig}).  These
findings are in agreement with the spatial localisation of functional
communities often found in brain networks~\cite{Sporns2015}, as well
as the hierarchical modularity exhibited by the \CE~connectome as
reported in Ref.~\cite{Bassett2010}. We remark that our
community detection method does not enforce a hierarchical
agglomeration of communities: the observed quasi-hierarchy and
spatial localisation is an intrinsic feature of the
\CE~connectome. In Fig.~\ref{S2_Fig} we quantify the deviation
of the community structure from a strict hierarchy.

At long Markov times, we find robust partitions containing 6 to 2
communities, denoted $\mathcal{A}$ to $\mathcal{E}$ in
Fig.~\ref{fig:Stability}.  Partition $\mathcal A$ comprises six
communities of varying sizes (from 9 to 104 neurons), well localised
along the body of the worm, as seen in Fig.~\ref{fig:Stability_soma}
(c.f. Section~2.2 in \url{www.wormatlas.org/neuronalwiring.html}).
The two large communities ($\mathcal{A}1$ and $\mathcal{A}2$) have head ganglia neurons of all
three functional types (S, I, M). In particular, $\mathcal{A}1$ contains ring
motor neurons and interneurons as well as the posterior neurons ALN
and PLN, whereas $\mathcal{A}2$ specifically gathers amphid neurons (e.g., AWAL/R,
ASKL/R, ASIL/R, AIYL/R) which feature prominently in the navigation
circuit responsible for exploratory behaviour~\cite{gray2005}.
Communities $\mathcal{A}3$, $\mathcal{A}4$ and $\mathcal{A}5$ in Partition $\mathcal A$ consist
predominantly of ventral cord motor neurons, differentiated by their
soma position along the body (Fig.~\ref{fig:Stability_soma}): $\mathcal{A}3$
contains frontal motor neurons (e.g. VD1 to VD3); $\mathcal{A}4$ consists of
mid-body motor neurons (e.g. VD4 to VD8); $\mathcal{A}5$ comprises posterior motor
neurons (e.g. VD9 and VD10).  Such partitioning is consistent with the
motor neuron segmentation model proposed for \CE~in
Ref.~\cite{Haspel2001}.  Finally, $\mathcal{A}6$ contains highly central neurons
such as AVAL/R or PVCL/R, which have been found to belong to a
\textit{rich-club}~\cite{towlson2013}, as well as interneurons linked
to mechanosensation and tap withdrawal functional
circuits~\cite{pan2010}.

The coarser Partitions $\mathcal B$ and $\mathcal C$ are
quasi-hierarchical merges of $\mathcal A$
(Figs.~\ref{fig:Stability}~and~\ref{fig:Stability_soma}).  For
instance, Partition $\mathcal C$ has three groupings: head ganglia
(merged $\mathcal{A}1$ and $\mathcal{A}2$), frontal motor neurons (merged $\mathcal{A}3$ and $\mathcal{A}4$), and a
tail subgroup (merged $\mathcal{A}5$ and $\mathcal{A}6$).  Interestingly, at later Markov
times, we obtain the distinct, coarser 3-community Partition $\mathcal
D$, which exemplifies how our method does not enforce a strict
hierarchy in the multiscale structure.  The three groups in Partition
$\mathcal D$ include a notable community of only three nodes
(interneurons AVFL/R and AVHR), which appear as a cohesive group only
at this particular timescale.  Prominent functional roles of AVF and
AVH neurons have been noted previously~\cite{varshney2011,Schaub2014}:
both AVF neurons are responsible for coordination of egg-laying and
locomotion~\cite{hardaker2001}.  In addition, spectral analyses of the
gap-junction Laplacian have shown that AVF, AVH, PHB and C-type motor
neurons are strongly coupled~\cite{varshney2011}.  Finally, the two
communities in the coarsest Partition $\mathcal E$ split the
connectome anatomically into a group with head and tail ganglia (red),
and another group predominantly with motor neurons (cyan).

\subsection{The effect of single and double neuron ablations on
  flow-based communities}
 
Laser ablation experiments are invaluable to probe the functional role
of neurons~\cite{chalfie1985,li2011,wakabayashi2004}, but are time
consuming and technically challenging.  We have used our computational
framework to assess the effect that an ablation of a single neuron, or
of a pair of neurons, has on the signal flow in the connectome.  To
this end, we compare the flow-based partitions obtained for the
ablated connectome against the original network.  If an ablation
creates large distortions in the flow structure, the partitions of the
ablated network will change drastically or become less robust compared
to those found in the unablated network.  We have carried out a
systematic computational analysis of \textit{all} single and double
neuron ablations in the connectome.

\subsubsection{Single ablations: disrupting the robustness and make-up of partitions}

\begin{figure}[htb!]
\includegraphics{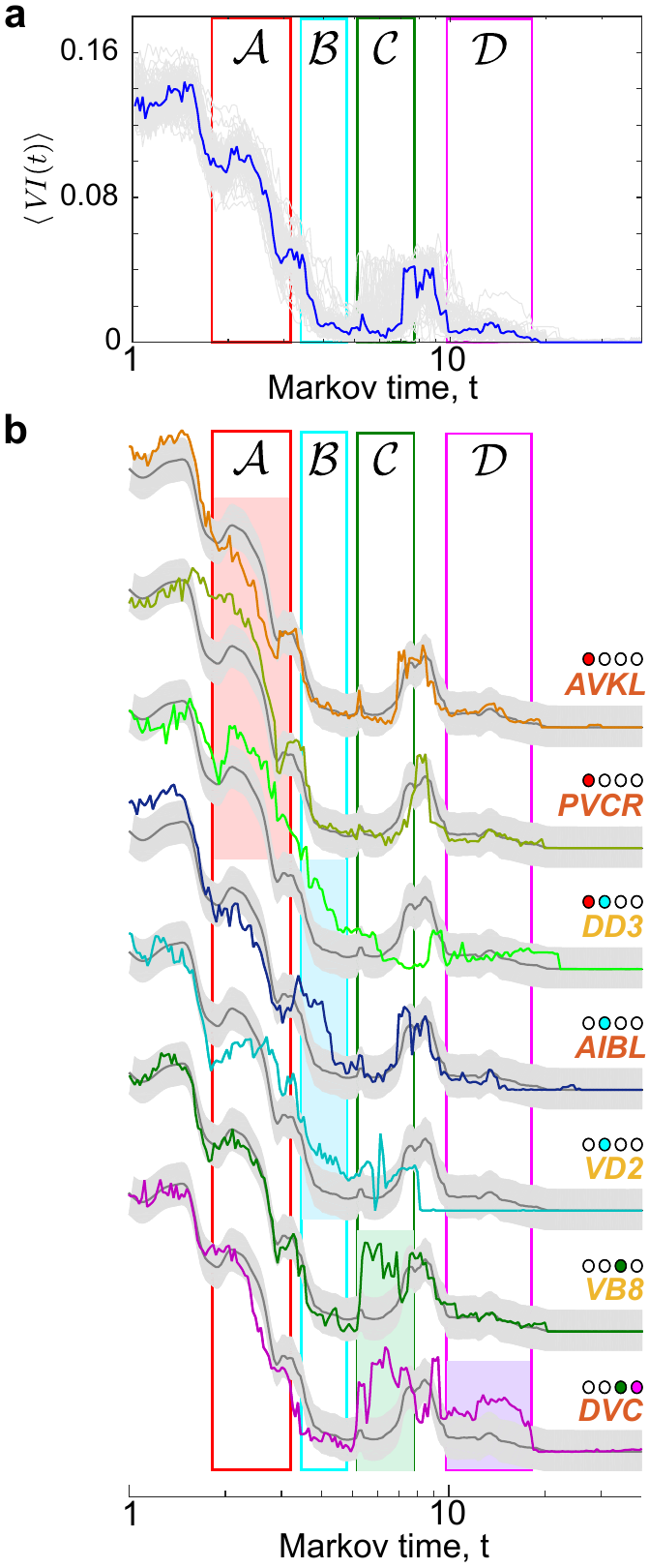}
\caption{\textbf{Single ablations that alter the robustness of
    partitions.}  \textbf{(a)} Ensemble of \VIt profiles of
  \textit{all} single node ablations (light gray lines) and the
  unablated connectome (blue). A Gaussian process (GP) is fitted to
  the ensemble of single ablations.  \textbf{(b)} The GP is described
  by the mean $\mu(t)$ (dark grey line) and standard deviation (grey
  bands). Sustained outliers from the GP are identified using a
  statistical criterion to find seven ablations that affect the
  different partitions, as indicated by the coloured dots. }
\label{fig:GPR}
\end{figure}

\paragraph{\textbf{Ablations that alter the robustness of partitions:}}
To find ablations that have a strong effect on the robustness of
Partitions $\mathcal A$--$\mathcal E$, we detect node deletions that
induce sustained changes in the robustness \VIt, i.e., they appear as
outliers with respect to a Gaussian Process fitted to the \VIt of the
ensemble of \textit{all} single node ablations (Fig.~\ref{fig:GPR}).
For details, see Section~`\ref{sec:deletion}'.

Only seven single ablations satisfy our criterion for a major
disruption of any of the Partitions $\mathcal A$--$\mathcal E$
(Fig.~\ref{fig:GPR}b). The ablations of interneuron PVCR or of the
motor neuron DD3 both decrease the robustness of Partition $\mathcal
A$.  Interestingly, PVCR ($\mathcal{A}6$) and DD3 ($\mathcal{A}4$) receive many incoming
connections from their own community.  Furthermore, both of these
neurons are critical for motor action: PVCR drives motion whereas DD3
coordinates it.  Another important ablation is that of interneuron
AVKL, which links community $\mathcal{A}1$ (head) with community $\mathcal{A}3$ (ventral cord)
and community $\mathcal{A}6$ (rich club).  The increased robustness of the
community structure upon ablation of AVKL would indicate a decreased
communication between these groups. The function of AVKL is uncharted
at present~\cite{WormAtlas}, suggesting further {\it in vivo} experimental investigations 
to explore any behavioural changes as a result of its ablation.

There are three important ablations in Partition $\mathcal B$: 
DD3 (again), VD2 (another D-type motor neuron yet on
the ventral side), and AIBL, an amphid interneuron.  
AIBL acts as a bridge between communities $\mathcal{A}1$ and
$\mathcal{A}2$, which merge in Partition $\mathcal B$
(Fig.~\ref{fig:Stability_soma}).  The prominent role of other amphid
interneurons will become apparent in the double ablations studied in
the next section.

Partition $\mathcal C$ is rendered non robust by the ablations of VB8
(a motor neuron responsible for forward locomotion) or of interneuron
DVC, with are both in community $\mathcal{A}5$.  DVC has links with communities
$\mathcal{A}3$, $\mathcal{A}4$, $\mathcal{A}5$ and $\mathcal{A}6$; hence its ablation affects the subsequent merging
of these groups.  Note that the ablation of DVC reduces the robustness
of both 3-way Partitions $\mathcal C$ and $\mathcal D$, thus blurring
the spatial organisation of motor neurons.  This indicates that DVC
might integrate feedback from different parts of the body, in
accordance with the fact that it has the highest number of gap
junctions in the connectome, as well as substantial chemical
synapses~\cite{Altun2009}.  

Our study of ablations that affect the robustness of partitions 
can be linked to the study of `community roles'~\cite{Guimera2005}. 
Using such categorisation, the neurons mentioned
above are classified as either connector or
provincial hubs (e.g., DVC is a `non-hub' connector node)~\cite{pan2010}. 

\paragraph{\textbf{Ablations that alter the make-up of the optimal partitions:}}

\begin{figure*}[tb!]
\includegraphics[width = 0.9\textwidth]{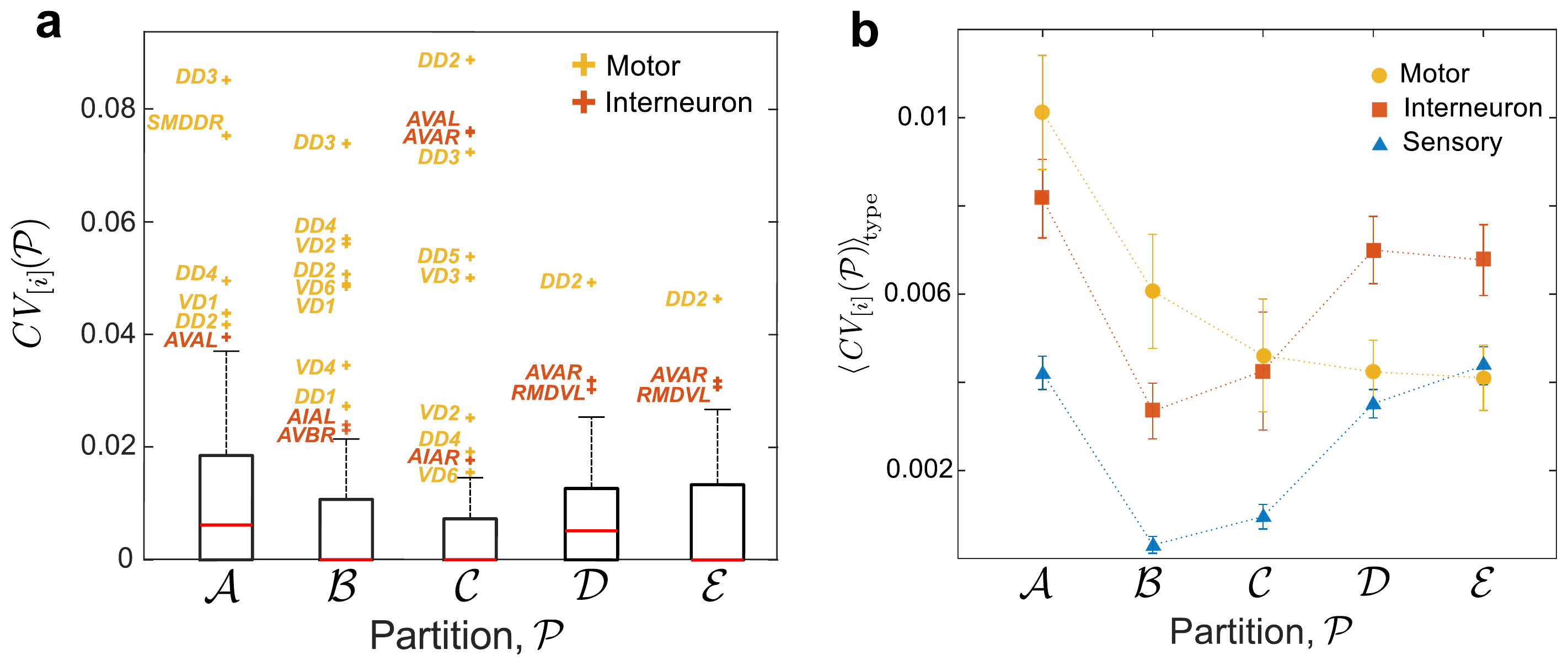}
\caption{ \textbf{Effect of single ablations on the make-up of
    different partitions as measured by the community variation.}
  (\textbf{a}) The disruption of every single mutation with respect to
  Partitions $\mathcal A$-$\mathcal E$ is quantified through
  $\textit{CV}_{[i]}(\mathcal{P})$, as defined in Eq.~\eqref{eq:CV}.
  The distribution of $\textit{CV}_{[i]}$ is represented by its median
  (red line) and the inter-percentile range (IPR) between the 10th and
  90th percentiles (box). The whiskers correspond to the IPR for each
  partition, and the single ablations detected as outliers are
  labelled.  (\textbf{b}) Effect of the single ablations
  $\textit{CV}_{[i]}(\mathcal{P})$ for each partition averaged over
  each type: sensory (blue), inter- (red) and motor neurons
  (yellow). On average, single ablations of motor neurons induce
  larger changes on the finer Partitions $\mathcal A$ and $\mathcal
  B$, whereas ablations of interneurons have a larger effect on the
  coarser Partitions $\mathcal D$ and $\mathcal E$. The error bars are
  the standard error of the mean. }
  \label{fig:CV_SIM} 
\end{figure*}

To measure how much the make-up of a partition is affected by an
ablation, we use the community variation \CV, defined in
Eq.~\eqref{eq:CV}.  A high value of $\textit{CV}_{[i]}(\mathcal{P})$
indicates a large disruption in partition $\mathcal{P}$ under the
ablation of neuron $i$.  Figure~\ref{fig:CV_SIM} shows the single
ablations with high \CV~with respect to Partitions $\mathcal
A$-$\mathcal E$, as detected through a statistical criterion based on
interpercentile ranges (see Section~`\ref{sec:deletion}').
Interestingly, none is a sensory neuron, indicating that the ablation
of sensory neurons is not influential for global flow at medium to
coarse scales, although they can have strong local effect on the
propagation of a particular stimulus.

Certain ablations are completely destructive of Partitions $\mathcal
A$ and $\mathcal B$.  In particular, the ablations of DD3 or SMDDR
induce severe changes in the network flow, so that no partition
similar to $\mathcal A$ is found at any Markov time.  In general,
ablations of D-type motor neurons coordinating motion (e.g. DD2, DD3,
VD1, VD2) have particularly severe effects for the medium resolution
Partitions $\mathcal A$ and $\mathcal B$.  Interestingly, D-type motor
neurons have significantly higher PageRank (median $0.0092$ compared
to median of $0.0018$ in the network; $p=1.7 \times 10^{-7}$;
one-sided exact test), and their synapses are critically embedded
edges with few alternative routes~\cite{Schaub2014}. Note that,
although robustness and make-up of partitions reflect different
effects, the ablation of motor neurons DD3 and VD2 substantially
alters both (see Figs.~\ref{fig:GPR} and~\ref{fig:CV_SIM}a).  In
addition, the ablation of any of the command neurons AVAR/L has
important effects on Partition $\mathcal C$. AVAR/L are highly central
neurons (with the highest in- and out- degree in the connectome) and
our method confirms that their ablation introduces heavy distortions
in the global flow of the connectome.  Finally, we observe that the
coarsest partitions $\mathcal D$ and $\mathcal E$ are strongly
perturbed upon ablation of ring motor neuron RMDVL. Experiments have
shown that ablating any of the RMD neurons diminishes the
head-withdrawal reflex~\cite{celegans2}.

Further confirmation of the importance of inter- and motor neurons is
given in Figure~\ref{fig:CV_SIM}b, where we show the \CV~of single
ablations averaged over the three types (S, I, M).  On average, motor
neurons tend to have a stronger effect on local organisation due to
their localised connectivity; this is reflected by the high \CV~in the
finer Partitions $\mathcal A$ and $\mathcal B$.  On the other hand, 
interneurons, which are mediators of information flow from sensory 
to motor neurons, can induce large changes in global flows, 
as shown by larger \CV~for the 
coarser Partitions $\mathcal D$ and $\mathcal E$.

\subsubsection{Double ablations: beyond additive effects}

We have also performed an exhaustive \textit{in silico} exploration of
all possible 38781 two-neuron ablations.  Specifically, we look for
synergistic pairs of neurons, i.e. pairs whose simultaneous ablation
induces supra-additive disruption.  To this end, we compare the
\CV~for each double ablation to the averaged \CV~of the corresponding
two single ablations, and use Quantile Regression to identify double
ablations with a combined effect significantly beyond the merely
additive (see Section~`\ref{sec:qr}').

\begin{figure}[t!]
   \includegraphics[width=0.45\textwidth]{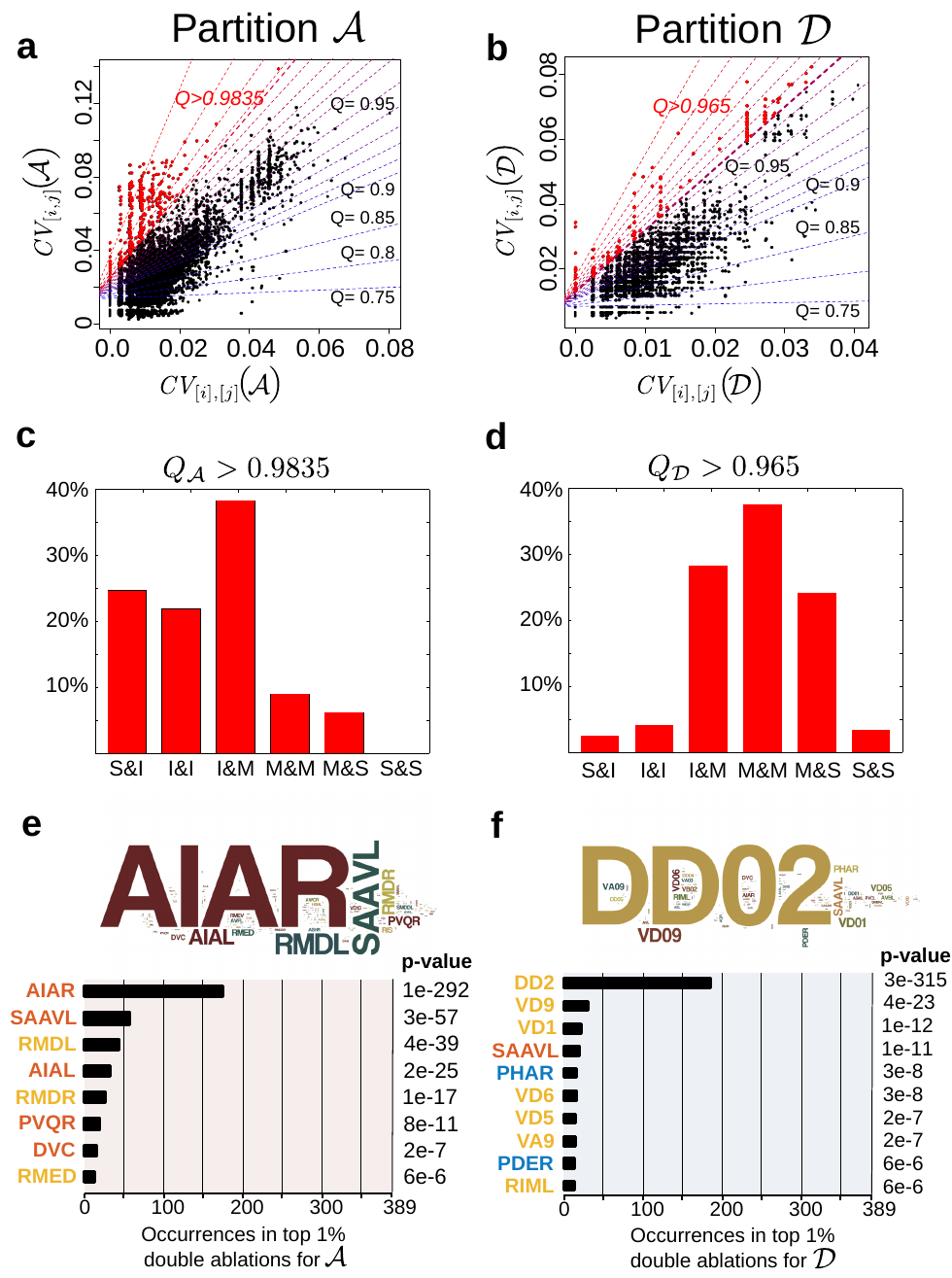}
   \caption{\textbf{Supra-additive double ablations.}  The combined
     effect of each two neuron ablation is compared against the
     additive effect of the corresponding two single ablations. The
     results of Quantile Regression of \CV~of the pair against the
     averaged \CV~of the two single ablations (see
     Section~`\ref{sec:qr}') are shown for: ({\bf a}) Partition
     $\mathcal A$ and (\textbf{b}) Partition $\mathcal D$.  The top
     1\% pairs with the largest supra-additive effect are found above
     the quantile scores $Q_\mathcal{A} > 0.9835$ and $Q_\mathcal{D} >
     0.965$, respectively.  These top 1\% double ablations are
     dominated by: {\bf (c)} interneurons for $\mathcal A$; {\bf (d)}
     motor neurons for $\mathcal D$.  Overrepresentation of neurons in the
     top 1\% supra-additive pairs for {\bf (e)} Partition $\mathcal
     A$ and {\bf (f)} Partition $\mathcal D$ was calculated using a one-sided Fisher exact test (unadjusted p-values are reported, and also provided for all neurons in section~\ref{S1_Data}).  
     Neurons with $p <10^{-5}$ are listed and the names of neurons are coloured 
     according to their type: S (blue), I (red), M (yellow). The word clouds are a
     visualisation of these over-representations. Computing a Bayesian quantile of higher prevalence of these neurons among the top 1\% pairs also supports these findings~\cite{gelman2014bayesian}.
}
  \label{fig:QR}
\end{figure}

We focus on disruptions to Partitions $\mathcal A$ and $\mathcal D$,
as prototypical of the medium and coarse resolutions, respectively
(Figure~\ref{fig:QR}).  We select the top 1\% of ablations for each
partition according to their supra-additive effect. Interestingly,
85\% of the top supra-additive double ablations for Partition
$\mathcal A$ contain at least one interneuron, whereas 90\% of the top
supra-additive double ablations for Partition $\mathcal D$ contain at
least one motor neuron (Fig.~\ref{fig:QR}c-d).  This observation
complements the results for single ablations in
Figure~\ref{fig:CV_SIM}.  For Partition~$\mathcal A$, maximal impact
of a single ablation is achieved through the deletion of motor
neurons, but double ablations containing interneurons are more
synergistic.  For the coarser Partition~$\mathcal D$, the most
disruptive single ablations are those of interneurons, yet on average
the most synergistically disruptive double ablations include motor
neurons.  Such joint effects underline the structured complexity of
the connectome network and reinforce the fundamental importance of I
and M neurons in the disruption of flows.  In particular, the relative
abundances of particular neurons in the top supra-additive pairs
(Fig.~\ref{fig:QR}e-f) show that interneurons AIAR/L, SAAVL and PVQR
and motor neurons RMDL/R are overly represented for Partition
$\mathcal A$. These neurons thus have a magnifying disruptive effect
for the medium scales of the connectome.  For the coarser Partition
$\mathcal D$, this magnifying effect on larger scales is induced
mostly by motor neurons DD2, VD9, VD1 and interneuron SAAVL.

If we consider the effect on both medium and large scales, only nine
double ablations appear in the top 1\% for both partition $\mathcal A$
and $\mathcal D$ (Table~\ref{tab:3}). Interestingly, none of these
pairs is linked by an edge in the connectome. Note that eight out of
these nine pairs contain interneuron AIAR. The amphid interneurons AIA
(along with AIB, AIY and AIZ) have a specific position in the
connectome: they receive synapses from sensory neurons driving motion.
Their prominent role in locomotion integration has been previously
discussed and backed by \textit{in vivo} ablation
experiments~\cite{wakabayashi2004}.  Our results indicate that the
deletion of pairs of neurons involving AIAR would have a particularly
magnifying effect on the disruption of the flow organisation at all
scales in the connectome. Note that the effect of AIAL in double ablations
is much less prominent. The asymmetry observed in how the ablations
of AIAR and AIAL affect the flows in the connectome is worth of
further experimental investigation.  The full set of outcomes of both
single and double ablations are presented in~\ref{S1_Data} as
a guide for possible experimental investigations.  

\begin{table}[ht!]
\centering
\caption{Double ablations within the top 1\% of supra-additive pairs 
  for both Partition $\mathcal A$ and $\mathcal D$ chosen according 
  to their quantile scores ($Q_\mathcal{A} > 0.9835$ 
  and $Q_\mathcal{D} > 0.965$).}
\label{tab:3}
\begin{tabular}{lccccc}
    \toprule
    Double ablation & & Neuron types & $Q_\mathcal{A}$ & & $Q_\mathcal{D}$ \\
    \cline{1-6}
    AIAR +  AQR   & & I \& S & 0.9975 & &0.9785 \\
    AIAR   +  AVEL   & & I \& I & 0.9975 & &0.9650  \\
    AIAR   +  DA2   & &I \& M & 0.9875 & &0.9785\\
    AIAR   +  VA2   & &I \& M & 0.9945 & &0.9655\\
    AIAR   +  VB2   & &I \& M & 0.9970 & &0.9785\\
    AIAR   +  VD5   & &I \& M &   0.9950 & & 0.9785\\
    AIAR   +  VD6   & &I \& M & 0.9950 &   &  0.9785\\
    AIAR   +  PVCR   & & I \& I & 0.9980 &  &  0.9785\\
    SAAVL + AQR  & & I \& S & 0.9955 & &0.9730 \\
    \botrule
\end{tabular}
\end{table}

\subsection{Identifying flow profiles in the directed connectome}

A complementary analysis of the directed connectome of \CE~is provided
by the Role Based Similarity (RBS) framework~\cite{Cooper2010,Cooper2010a}, 
which identifies groups of nodes with similar \textit{flow profiles} in the network 
without imposing \textit{a priori} the type or number of groups. 
Such groups of neurons display the same character (or \textit{flow role}) 
in terms of their role in the generation, distribution and consumption of
  flow in the network. Briefly, RBS obtains a {\it flow profile} for each node 
  from its incoming and outgoing flows at all scales. We then group the nodes into
  classes (`flow roles') with similar in- and out-flow patterns.  
  Because they include information at all scales, flow roles capture 
  nuanced information about the network, beyond pre-defined categories 
  (e.g., sources, sinks, hubs) or combinatorial notions based on immediate 
  neighbourhoods (e.g., roles from Structural Equivalence~\cite{Lorrain1971} 
  and Regular Equivalence~\cite{Everett1994}).  Details of the RBS methodology are
given in Refs.~\cite{Cooper2010, Cooper2010a, Beguerisse2013,
  Beguerisse2014}, and summarised in Section~`\ref{sec:RBS}' and in Fig.~\ref{S4_Fig}.

\begin{figure*}[h!]
    \includegraphics{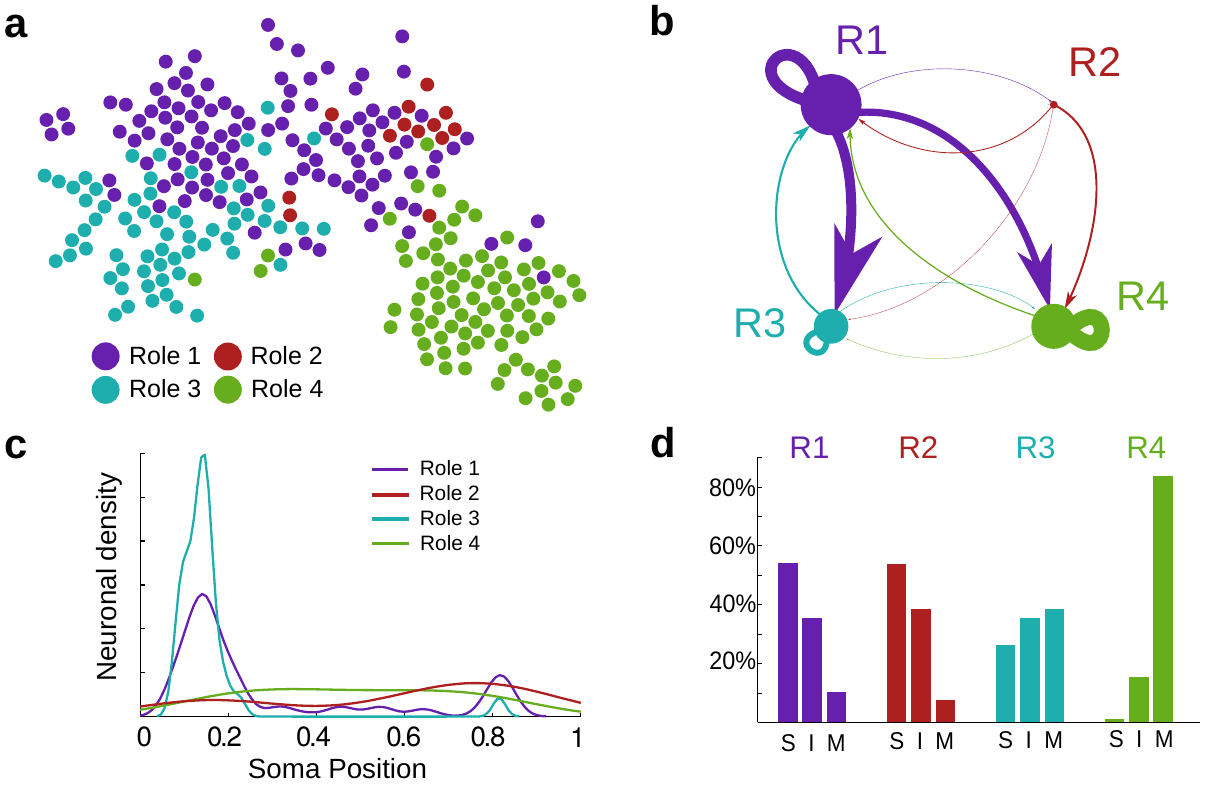}
    \caption{\textbf{Flow roles for neurons in the \CE~connectome.}
      ({\bf a}) Using RBS, we detect four flow roles in the directed
      connectome.  ({\bf b}) The
      coarse-grained representation summarises the flow profiles of
      the roles: two upstream roles (R1, R2), with a dominant source
      character and high PageRank (Fig.~\ref{S6_Fig}), and two
      downstream roles (R3, R4), with a dominant sink character and
      lower PageRank.  Yet each role has distinctive in- and out-flow
      patterns in relation to the others.  ({\bf c}) Spatial density
      of neurons for each flow role represented as a function of 
      the normalised soma position: R1 and R3 are localised
      predominantly in the head; R2 and R4 are spread out along the
      body.  Note how the upstream flow role R2 has noticeable localisation
      in the tail.  ({\bf d}) The percentages of sensory (S), inter-
      (I) and motor neurons (M) in each role underline their
      functional differences.  }
  \label{fig:RBS}
\end{figure*}

In the \CE~connectome, we identify four distinct classes of neurons
according to their flow profiles (Fig.~\ref{fig:RBS}). These flow roles
are distinct from the groupings into communities (see an analysis of communities and
their mix of flow roles in Fig.~\ref{S5_Fig}). Two of the roles (R1 and R2) have a
dominant `source' character (i.e., higher average in-degree than
out-degree) and contain most of the nodes with high PageRank (Fig.~\ref{S6_Fig}). The other two roles (R3 and R4) have a dominant `sink'
character and nodes with low PageRank. Note, however, that these roles
are not just defined by average properties, but by their global
flow patterns in the network.  As seen in Fig.~\ref{fig:RBS}b, R1 is
upstream from R3 and R4, whereas R2 is mostly upstream from R4.
Furthermore, R4 is an almost pure downstream module, whereas R3 has a
stronger feedback connection with R1.

The RBS flow roles are linked to physiological
properties of the neurons (Fig.~\ref{fig:RBS}c-d).  R4 corresponds to
a group of motor neurons (mostly ventral chord motor neurons)
consistent with its downstream character, whereas R1 is a group of
mostly sensory and inter-neurons with heavy localisation in the head.
R3 is a group with a balanced representation of all three types of
neurons (including some polymodal neurons) localised in the head.
Indeed, most ring neurons in R3 are in community $\mathcal{A}1$, indicating a
self-contained unit that process head-specific behaviour, such as
foraging movements and the head withdrawal reflex~\cite{WormAtlas}.

Our RBS analysis also reveals a specific flow profile (R2) containing 13
neurons (mainly sensory and interneurons, mostly upstream from the
motor neurons in R4), the majority of which are responsible for escape reflexes 
triggered in the presence of noxious factors (Table~\ref{tab:escape}).
This group  can be seen as a group of \textit{escape response neurons}
and include: the PVDL/R
neurons, which sense cold temperatures and harsh touch along the body;
FLPL/R, which perform the equivalent task for the anterior region; PHB
neurons responsible for chemorepulsion; PHCR, which detects noxiously
high temperatures in the tail; SDQL and PQR, which mediate high oxygen
and CO$_2$ avoidance, respectively; and PLMR, a touch mechanosensor in
the tail~\cite{white1986}.  This escape response group is heavily
over-connected to command neurons AVAL/R, AVDL/R, DVA, PVCL/R, all of
which modulate the locomotion of the worm.  (Specifically, there are
48 connections from R2 to these particular command neurons in contrast
to the $\sim$12 connections expected at random.)  Note that AVDL/R and
DVA are in R1, whereas AVAL/R and PVCL/R are in R4; the R2 group thus
links directly to motor locomotion neurons across the worm.  We remark
that this group of neurons was found exclusively through the analysis
of their all-scale in/out flow profiles, without any other extrinsic
information.

\begin{table}[tbh!]
  \caption{\textbf{Role 2 (R2) neurons:} The thirteen neurons
    identified in R2 constitute a group of \textit{escape response
      neurons} containing mostly sensory and inter-neurons linked with
    escape reflex reactions in response to different noxious stimuli.}
  \begin{tabular}{lll}
    \toprule
    {\bf Neuron(s)} && {\bf Noxious factor} \\
    \hline
    FLPL/R & &Harsh touch, low temperature (head) \\
    PHBL/R & & Chemicals \\
    PHCR & &High temperature (tail) \\
    PLMR & & Gentle touch (tail) \\
    PQR & & CO$_2$ \\ 
    PVDL/R & & Harsh touch, low temperature \\
    SDQL & & High O$_2$ \\ 
    SAAVL/R & & No known factor\\
    VD11 & & No known factor\\ 
    \botrule
  \end{tabular}
  \label{tab:escape}
\end{table}

\subsection{Information propagation in the connectome:
  biological input scenarios}
\label{sec:signal}

Despite its modest size, the nervous system of \CE~can sense and react
to a wide range of mechanical, chemical and thermal
factors~\cite{WormAtlas}.  Standard notions in neuroscience hold that
stimuli lead to motor action due to information progressing from
sensory through inter- to motor neurons~\cite{Jarrell2012}. However,
the underlying mechanisms and precise signal flows are still far from
understood.  In the absence of measurements probing such pathways, and
as a first approximation to more realistic nonlinear dynamical models,
we use here simplified diffusive dynamics (see Section
`\ref{sec:diffusion}') to mimic signal propagation in the \CE~directed
network.  Such an approach, already suggested by Varshney {\it et
  al}.~\cite{varshney2011}, is naturally linked to MS multiscale
community detection and to the identification of RBS flow roles, since
both Markov Stability and Role Based Similarity are intrinsically
defined in terms of a diffusive process on the graph.

To mimic the propagation of stimuli associated with particular
biological scenarios, a normalised initial flow vector
$\boldsymbol{\phi}(0)$ is localised at specific input neurons and we
observe the decay towards stationarity under
Eq.~\eqref{eq:heatsoldir}:
\begin{align}
    \boldsymbol{\theta}(t)=\boldsymbol{\phi}(t)-\boldsymbol{\pi}.
\end{align}
 We also define $\boldsymbol{q}(t)$, which will be used to
detect overshooting neurons:
\begin{equation}
    q_i(t) = \frac{\phi_i(t)}{\pi_i} = 1 + \frac{\theta_i(t)}{\pi_i}.
\end{equation}
Initially, $\theta_i(0)$ is positive only for the input neurons
where we inject the signal, and negative for all other neurons. 
Asymptotically, the vector of flows $\boldsymbol{\phi}(t)$
approaches the stationary solution $\boldsymbol{\pi}$, and
$\theta_i(t) \rightarrow 0, \forall i.$ 
However the approach to the stationary value can be qualitatively different.
In some cases, $\theta_i(t)$ can become positive, if neuron $i$ receives an influx of
flow that drives it to `overshoot' above its stationary value; 
in other cases, neurons approach stationarity without overshooting. 
The different behaviour depends on the particular initial input and the relative 
location of each neuron in the network.

Motivated by several experimental studies, we have conducted four case studies 
corresponding to different biological scenarios in which the input is localised 
on specific neurons:
\begin{enumerate}
\item[(i1)] Posterior (tail) mechanosensory
  stimulus~\cite{li2011,chalfie1985}: PLML/R, PVDL/R, PDEL/R
\item[(i2)] Anterior (head) mechanosensory
  stimulus~\cite{li2011,chalfie1985}: ADEL/R, ALML/R, AQR, AVM,
  BDUR/L, FLPL/R, SIADL/R
\item[(i3)] Posterior (tail) chemosensory stimulus (also reported as anus
  mechanosensory stimulus)~\cite{hillard2002,li2011}: PHAL/R, PHB/R
\item[(i4)] Anterior (head) chemosensory stimulus~\cite{hillard2002}: ADLL/R,
  ASHL/R, ASKL/R.
\end{enumerate}

\begin{figure*}[tb!]
    \includegraphics[width=0.9\textwidth]{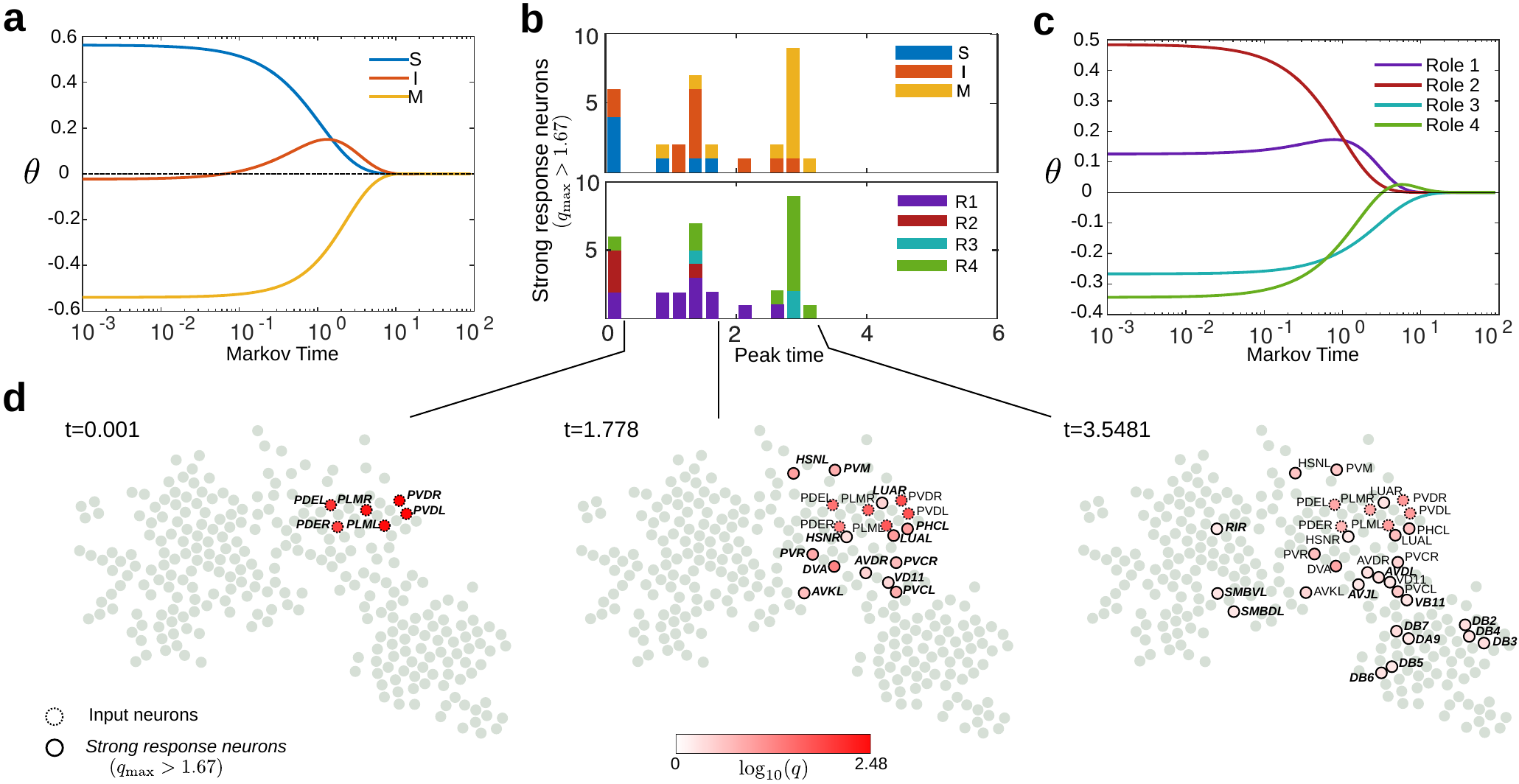}
    \caption{\textbf{Signal propagation of posterior mechanosensory
        stimulus (i1).}  \textbf{(a)} As stationarity is approached
      ($\boldsymbol{\theta}(t) \to 0$), the input propagates from
      sensory to motor neurons through an intermediate stage when
      interneurons overshoot.  \textbf{(b)} Signal propagation as a
      cascade of strong response neurons (32 neurons with
      $q_\text{max,i}> 1+2/3$) with peak times concentrated around two
      bursts. The number of neurons are colored according to type
      (top) and role (bottom).  Note the overall trend S~$\to$
      I~$\to$~M during the propagation of strong responses, and how
      the sequence of strong response neurons also reflects the
      connectivity between roles propagating roughly from R2 to R1 and
      finally to R3.  \textbf{(c)} The input (i1), which is highly
      localised on R2 neurons, diffuses quickly to R1 neurons and
      induces an overshoot of R4 neurons followed by slower diffusion
      into R3 neurons.  \textbf{(d)} Stages of signal propagation in
      the network showing the strong response neurons that have peaked
      at each time.  }
    \label{fig:Signal}
\end{figure*}

We exemplify the procedure in detail through the posterior
mechanosensory stimulus (i1), but detailed results for the other
stimuli are provided in the Fig.~\ref{S8_Fig}, Fig.~\ref{S9_Fig}, and Fig.~\ref{S10_Fig}.  As shown in
Figure~\ref{fig:Signal}a, the signal proceeds `downstream' following
the expected biological information processing sequence,
S$\to$I$\to$M.  The signal is initially concentrated on the input
neurons (mostly sensory); then propagates out primarily to
interneurons, which overshoot and peak at $t\approx 1.5$; and is then
passed on to motor neurons, which slowly increase towards their
stationary value.

The flow roles obtained above provide further insight into the propagation of
stimuli.  As seen in Figure~\ref{fig:Signal}c, the input for the tail
mechanosensory scenario (i1) is heavily concentrated on R2 neurons
(the escape response group), from which the signal flows quickly
towards the other upstream (head) group R1, followed by propagation
towards the downstream group R4.  Finally, the signal spreads more
slowly to R3, the head-centric downstream unit.  This pattern of
propagation carries onto the sequence of strong response neurons
(Figure~\ref{fig:Signal}b), and reflects the fact that R2 contains
posterior upstream units, and mirrors the strong connectivity of R2
with motor neurons in R1 (AVDL/R and DVA) and R4 (PVCL/R), as
discussed above.

To detect key neurons comprising the specific propagation pathways, we find 
\textit{strong response neurons}, i.e., those with large overshoots
relative to their stationary value,
\begin{equation*}
q_{\text{max},i}  = \max_t  q_i(t) > 1+\frac{2}{3}. 
\end{equation*}
See Fig.~\ref{S7_Fig} for a full description of the procedure.
According to this criterion, we obtain 26 strong response neurons for scenario (i1). 
The neurons have large overshoots in two time windows after the inital input
(Figure~\ref{fig:Signal}b).  The details of the signal propagation
(Figure~\ref{fig:Signal}d) show that a first wave of peak responses
(around $t\approx 1$) corresponds mostly to overshooting interneurons,
including AVDL/R and DVA, responsible for mechanosensory integration,
and PVCL/R, drivers of forward motion~\cite{chalfie1985,WormAtlas}.
The second wave of peaks (around $t\approx 3$) contains predominantly
ventral B-type motor neurons, e.g., DB2-7 and VB11. Such B-type motor
neurons are responsible for forward motion. Hence the progression of
overshooting neurons suggests a plausible biological response for a
posterior mechanosensory stimulus~\cite{WormAtlas,li2011}. 
The overshooting behaviour of the neurons is not captured by other static
measures of the network (e.g., in/out degree or pagerank), as shown in~\ref{S12_Fig}.

\subsubsection{Comparison with other biological scenarios}
Detailed results of propagation under the other biological scenarios
(i2)-(i4) from the experimental literature are presented in Fig.~\ref{S8_Fig}, Fig.~\ref{S9_Fig}, and Fig.~\ref{S10_Fig}.  
The overall progression of the signal from S to I to M is observed with small
differences in all scenarios.  However, the different scenarios
exhibit distinctive participation of the flow roles. In particular,
both posterior stimuli (i1) and (i3) spread from R2 neurons quickly
into R1 neurons and R4 (motor) neurons, with weak propagation into R3
neurons .  On the other hand, anterior stimuli
(i2) and (i4) spread from the R1 group strongly into R3 neurons and
also quickly to R2 neurons, with only weak spreading into R4 neurons.
In cases (i1)-(i3) information flows fast out of R2 towards motor
neurons, as could be expected from neurons triggering an escape
response. Interestingly, the (i4) scenario does not feature any strong
response neurons in the R2 group.  

\begin{figure*}[tb!]
    \includegraphics{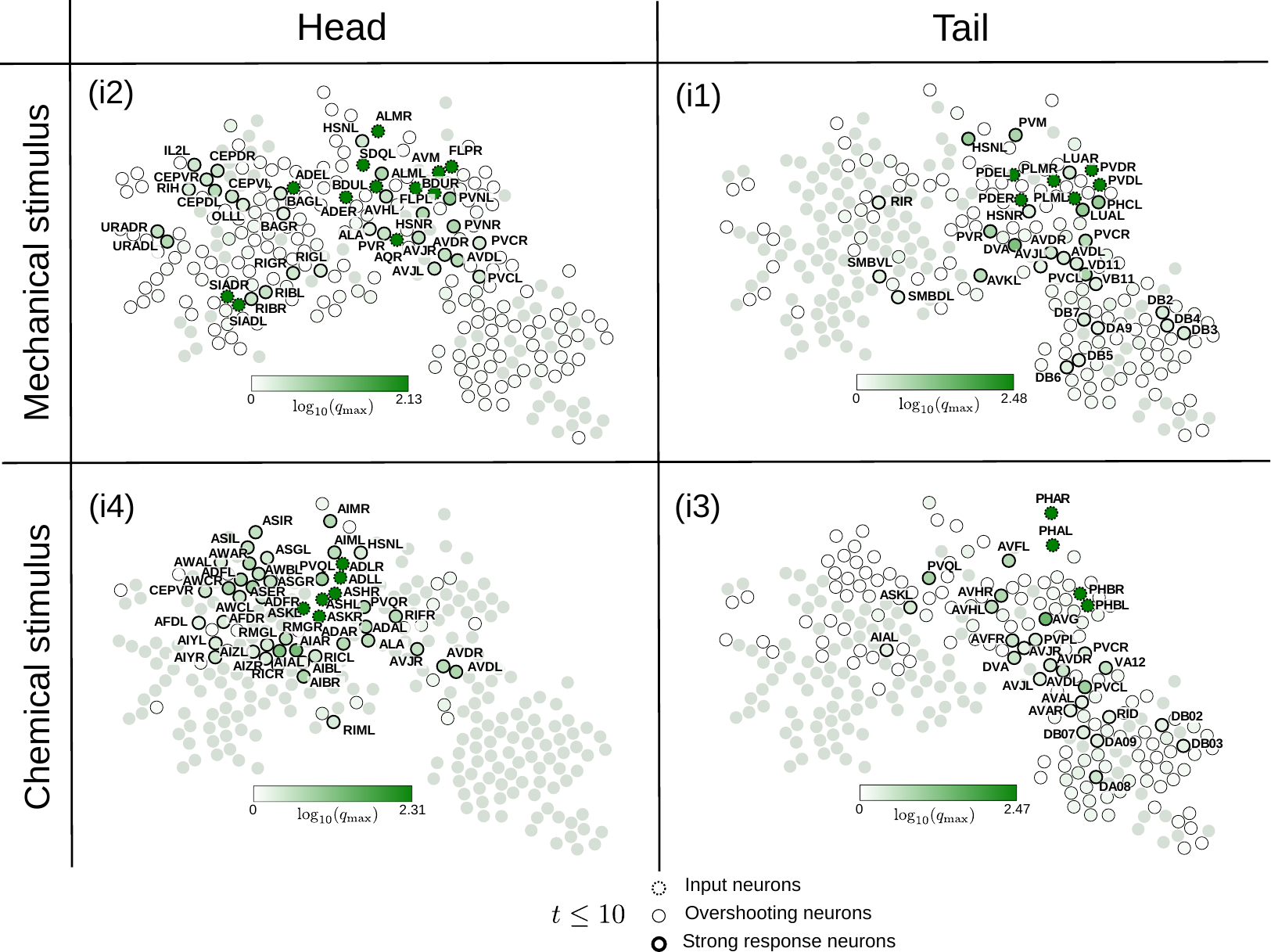}
    \caption{\textbf{Summary of signal propagation in the four
        biological scenarios.}  The specific pathways for the signal
      propagation for each of the scenarios (i1)-(i4) are shown,
      highlighting the input, strong response ($q_\text{max,i} > 5/3$) 
      and overshooting neurons ($q_\text{max,i} > 1$). 
      The input and strong response neurons are labelled for each biological scenario.}
    \label{fig:Scenarios}
\end{figure*}

As shown in Fig.~\ref{S8_Fig} - Fig.~\ref{S10_Fig}, and summarised in
Fig.~\ref{fig:Scenarios}, the signal propagation pathways have
distinctive characteristics for each of the scenarios.  For instance,
although the posterior chemosensory scenario (i3) shows strong
similarities to (i1) at earlier stages (input mostly R2 and strongly
responding interneurons PVCL/R, AVDL/R, AVJL, DVA), they show
differences in the motor neurons exhibiting a strong overshoot. 
In particular, for (i3) A-type neurons (DA8, DA9, VA12) responsible for backward motion
are present in addition to B-type neurons (DB2, DB3, DB7).

The anterior (head) scenarios (i2) and (i4)  inputs show a localised
propagation mostly in head-centric groups R1 and R3.
For the anterior mechanosensory scenario (i2), command
interneurons such as PVCL/R, AVDL/R respond strongly, together with
ring interneurons, such as RIGL/R and RIBL/R.  In this case, only
small excitation of ventral cord motor neurons is attained. Instead,
we observe strong responses of polymodal ring motor neurons, such as
URADL/R and SIADL/R, and of sensory neurons CEPVL/R and
CEPDL/R, even though these CEP neurons receive no external
input. Interestingly, CEP neurons are reported~\cite{WormAtlas} 
to be functionally redundant with nose touch receptors ADE, 
where the input signal is located.  
Upon anterior chemical stimulation (i4), a bulk of flow is captured within the 
neuronal ring and induces strong response from chemosensory neurons 
such as PVQL, ASKL, AWAL/R, AWABL/R
and AWACL/R, as well as interneurons RICL/R, RMGL/R, AIAL/R and
AIBL/R, which are specific for integrating chemo-sensation.  Indeed, several
of these neurons also appear in the posterior chemosensory stimulus (i3). 
A summary of the strong response and overshooting
neurons for all scenarios is presented in Figure~\ref{fig:Scenarios}.

\section{Discussion}
We have presented an integrated network-theoretic analysis of the
\CE~connectome~in terms of directed flows. We exploit the connection
between diffusive processes and graph-theoretical properties, which
intimately links structure and dynamics, to elucidate dynamically
relevant features in the connectome. Although diffusive processes are
a coarse approximation of physiological signal propagation, they can
be used to extract systemic dynamical features,
specifically in the case of non-spiking neuronal systems such as
\CE~\cite{varshney2011}.

Using the Markov Stability (MS) framework, we have identified
flow-based groupings of neurons in the \CE~connectome at different
levels of granularity.  Previous studies~\cite{pan2010, sohn2011,
  pavlovic2014} have aimed at uncovering modules based on structural
properties of the network, usually considering a particular scale so
as to find one partition (e.g., modularity at the standard
resolution). In section \ref{S1_Text} we provide a detailed comparison of MS
multiscale flow structures against partitions found by
modularity~\cite{pan2010,sohn2011}, stochastic block
models~\cite{pavlovic2014} and the MapEquation~\cite{Edler2015}.
The partitions obtained by MS at a particular scale are closer to those 
obtained with directed modularity. The MS framework, however, 
provides a multiscale description across all scales by sweeping the
Markov time~\cite{Schaub2012}, respecting and exploiting
directionality. In doing so, it reveals an intrinsic,
quasi-hierarchical organisation of the connectome, giving insight into
relevant features of signal propagation.  The partitions found by MS
are in good agreement with \CE~physiology, and
summarise previously observed features, such as
the hierarchical and spatial organisation of neuronal
communities~\cite{Bassett2010,Sporns2015}.
  
The obtained flow-based organisation highlights the
prominent position of particular neurons, such as AVF and AVH, and
allows for a systematic exploration of single and double ablations
most disruptive of signal flows, thus providing 
insight into candidate neurons for further experimental
investigations. Examples of such neurons 
include, among others: the synergistic effects caused by neuron AIAR
in double ablations; the global role of D-type motor neurons, which
often appear as relevant in single ablations;  or the role of 
polymodal (I/M) SAAVL/R head neurons~\cite{WormAtlas},
about which little is known but which appear in the R2 group and are
salient in our ablations. Several other examples are discussed in the text,
and further such hypotheses may be formulated based on the full set 
of ablation scores we provide in section~\ref{S1_Data}
as a resource to experimentalists investigating the
physiology of particular neurons.  

Other methods can be used to study the effect of ablations using,
for example, measures of centrality, efficiency or information 
transfer~\cite{Marinazzo2012,Marinazzo2014}. Our study of ablations
gives distinct results, as shown in Fig.~\ref{S3_Fig}. For instance,
because our measures focus on the disruption of the 
flow community structure at different scales, our approach can provide a structured view
of the effect of ablations for different neuron types, 
as shown in Figs.~\ref{fig:GPR}--\ref{fig:QR}.

As a complementary flow-based perspective, we have used Role Based Similarity to 
identify classes of neurons with similar patterns of flow in the \CE~nervous system.  
Rather than reflecting any measure of connectedness in the network, 
such \textit{flow roles} (or flow profiles) reflect similar roles in the 
generation, distribution and consumption of flow in the directed connectome. 
In previous work, neurons have been assigned to roles by exploring the core-periphery
structure~\cite{chatterjee2007}, or by examining the connections of
nodes within and between communities~\cite{Guimera2005,Klimm2014}.
Other notions of roles have been based on the use of centrality scores, or on
combinatorial notions of social neighbourhoods, as in regular and
structural equivalence~\cite{Everett1994,Lorrain1971}. 
RBS takes a different approach by grouping neurons according to
their patterns of in/out flows at multiple scales in the graph, irrespective
of their community membership and going beyond standard
classifications~\cite{Beguerisse2013,Beguerisse2014}. 
See \ref{S2_Text} and Fig.~\ref{S6_Fig} for a comparison of RBS flow roles, regular equivalence and community roles.

The RBS analysis of flow profiles finds two groups of mostly upstream neurons and 
two groups of mostly downstream neurons, yet
with a specific inter-connectivity pattern. 
In particular, the analysis singles out a small group of upstream neurons (R2), 
which is functionally related to escape responses from noxious factors, 
and could also be the object of further experimental investigation.
The RBS roles are also informative in conjunction with signal
propagation from `input-response' \textit{in silico}
biological scenarios (see Fig.~\ref{S11_Fig}).  
In particular, the R2 group plays an important role in posterior
biological stimuli, channelling stronger and faster responses, 
whereas R3 (the downstream, head-centric
group) constitutes a self-contained set of neurons mainly accessible via the
upstream, head-centric R1 group.  Therefore, the propagation profiles
obtained for different biological scenarios suggest a
graded organisation of the roles of nodes in terms of upstream-downstream
information, which could provide valuable insight into
functional circuits. 

Interesting theoretical extensions of the current work
would include considering the \CE~connectome
as a multiplex network; taking into account the different types of
synapse in a more explicit fashion; and enriching the dynamics of the
model by incorporating the effects of inhibitory synapses
and nonlinearities in the dynamics.
Furthermore, one may explore more intricate dynamics by incorporating
the memory of information flow using higher order Markov models~\cite{Rosvall2014,Salnikov2016}.

Our computational tools could be used in conjunction with
experimental techniques, as an aid to the generation of functional
hypotheses for experimental evaluation. With the eventual aim of
linking wiring properties of the connectome with information
processing and functional behaviour,
high throughput experiments (e.g., systematic ablation of several
neurons) coupled with advancements in neuronal monitoring 
that can allow recordings from thousands of neurons simultaneously~\cite{Ahrens2013}
could deliver time course measurements
to characterise signal propagation in relation to function.
Another interesting area of future work would be the evaluation
of ablation and propagation scenarios as related to 
quantitative behavioural investigations upon more general 
ablational/mutational strategies in \CE~\cite{Stephens2008,Yemini2013,Brown2013}, 
as well as comparative studies of the flow architecture in different 
nematode species~\cite{Bumbarger2013}.
Such comparative analyses between the functional and structural network 
of the connectome could yield valuable information 
in bridging the relation between structure and function in network neuroscience.

\section{Methods}
\subsection{The \textit{C. elegans} neuronal network}

The information of the large component of the connectome network is
encoded into the $n\times n$ adjacency matrix $A$ ($n=279$), where
entry $A_{ij}$ counts the total number of synapses (both chemical
synapses and gap junctions) connecting neuron $i$ to neuron
$j$~\cite{varshney2011}.  Note that chemical synapses are not
necessarily reciprocal, hence $A \neq A^T$.  Therefore the connectome
is a \textit{directed, weighted network}.  The network is relatively
sparse, with 2990 edges: 796 edges formed by gap junctions only; 1962
containing only chemical synapses; 232 edges with both gap junctions
and chemical synapses present.  The vector of out-strengths, which
compiles the sum of all synapses for each neuron, is
$\vec{d}=A\vec{1}$ (where $\vec{1}$ is the $n \times 1$ vector of
ones).  The average out-strength per neuron is 29; ranging from the
maximum (256) attained by neuron AVAL to the minimum (0) attained by
the motor neuron DD6, which is the only sink in the network.  The
network is {\it not} strongly connected.

\subsection{Propagation dynamics in the network}
\label{sec:diffusion}

Methods with different levels of complexity have been used to study
signal propagation in the \CE~connectome (see, e.g.,
Refs. \cite{zaslaver2015,koch1999,ferree1999,varshney2011,Jarrell2012}).
Here, we use a continuous-time diffusion process as a simple proxy for
the spread of information in this neuronal network.  Note that gap
junctions may be simply modelled as linear resistors and, although
chemical synapses are likely to introduce nonlinearities, their
sigmoidal transfer functions may be well approximated by a
linearisation around their operating point. Indeed, as remarked by
Varshney et al.~\cite{varshney2011}, such an approach has additional
merit in \CE, where neurons do not fire action potentials and have
chemical synapses that release neurotransmitters
tonically~\cite{Goodman1998}. Thus, linear systems analysis is in this case 
an appropriate tool that can provide valuable
insights~\cite{varshney2011}. Interestingly, athough simplified, such
linear models have been successfully applied even to the analysis of
spatio-temporal behaviour of strongly nonlinear neuronal
networks~\cite{Schaub2015}.

The signal on the nodes at time $t$ is represented by the $1 \times n$
row vector $\boldsymbol{\phi}(t)$ governed by the differential
equation
\begin{equation}
    \dfrac{d\boldsymbol{\phi}}{dt} = 
    \boldsymbol{\phi} \left[ M  -  I \right],
    \label{eq:heatdir}
\end{equation}
where $I $ is the identity matrix and $M $ is the
transition matrix defined as follows:
\begin{equation}
    M  = \tau D^\dagger{A} +
    \frac{1}{n}\left[\left(1-\tau\right)\vec{1} +
        \tau \,\boldsymbol{\mathbbm{1}}_{d_i = 0}\right] \vec{1}^T.
    \label{eq:mdef}
\end{equation}
Here, $\tau \in (0,1)$ is the Google teleportation parameter (and we
take $\tau = 0.85$ as is customary in the literature);
$\boldsymbol{\mathbbm{1}}_{d_i = 0}$ is the indicator vector of sink
nodes; and the diagonal matrix $D^\dagger$ is the pseudo-inverse of
the degree matrix:
\begin{equation*}
    D _{ii}^\dagger=\left\{\begin{array}{ll}
            0 & \mbox{if}~~d_i=0 \\
            1 / d_i &\mbox{if}~~d_i\neq 0. \\
        \end{array} \right.
\end{equation*}
The matrix $M$ describes a signal diffusion along the directed edges
with an additional re-injection of external `environmental noise':
each node receives inputs from its neighbours (which transmit flow
along their outgoing links according to their relative weight with
probability $\tau$) and receives a constant external re-injection of
size $(1-\tau)/n$.  For pure sinks, the outgoing flow is uniformly
redistributed to all nodes so as to avoid the signal accumulating at
nodes with no out-links.  Mathematically, this reinjection of
probability (known as teleportation in the networks literature)
guarantees the existence of a unique stationary solution for
Eq.~\eqref{eq:heatdir}, even when the network is not strongly
connected~\cite{Page1999, Lambiotte2014}. Biophysically, the
teleportation can be understood as modelling the random interactions
with the external environment.

Let $\boldsymbol{\phi}(0)$ be the input, i.e., the signal at $t=0$. 
The solution of Eq.~\eqref{eq:heatdir} is then:
\begin{equation}
    \boldsymbol{\phi}(t) = \boldsymbol{\phi}(0) \exp\left(t\,[M  - I ]\right),
    \label{eq:heatsoldir}
\end{equation}
with stationary solution $\boldsymbol{\phi}(t \rightarrow \infty)=
(\boldsymbol{\phi}(0) \cdot \mathbf{1}) \, \boldsymbol{\pi}$, where
$\boldsymbol{\pi}$ is the dominant left eigenvector of $M$, known as
PageRank~\cite{Page1999}.  Therefore, under a unit-normalised input,
$\boldsymbol{\phi}(t)\cdot\vec{1}=1$ $\forall\, t$, and
the stationary solution is~$\boldsymbol{\pi}$.  

\subsection{A dynamical perspective for community detection in graphs:
  Markov Stability}
\label{sec:stability}

The diffusive dynamics~\eqref{eq:heatdir} can be exploited to reveal
the multiscale organisation of the \CE~connectome using the Markov
Stability community detection
framework~\cite{Delvenne2010,Delvenne2013,Lambiotte2014}.  Markov
Stability finds communities across scales by optimising a cost
function related to this diffusion (parametrically dependent on time)
over the space of all partitions.

More formally, a partition $\mathcal{P}$ of the $n$ nodes of the
network into $m$ non-overlapping communities is encoded as a $n \times
m$ \textit{indicator matrix} $H_\mathcal{P}$:
\begin{equation} 
    {[H_\mathcal{P}]}_{ic}=\left\{
        \begin{array}{ll} 1 &  \mbox{if node $i$ belongs to community $c$} \\ 
            0 &   \mbox{otherwise.} 
        \end{array}
    \right.  \label{eq:Hdef} 
\end{equation} 
Given a partition matrix $H_\mathcal{P}$, we define the time-dependent
\textit{clustered autocovariance matrix}:
\begin{equation} 
\label{eq:R_def}
R(t, H_\mathcal{P}) =
    {H_\mathcal{P}}^T\left[ \Pi\exp(t[M  - I ])-
        \boldsymbol{\pi}\boldsymbol{\pi}^T \right] H_\mathcal{P}, 
\end{equation}
where $\Pi=\mathrm{diag}(\boldsymbol{\pi})$.  The matrix entry
${[R(t, H_\mathcal{P})]}_{c f}$ quantifies how likely it is that a random walker
starting in community $c$ will end in community $f$ at time $t$, minus
the probability for such an event to happen by chance.  To find groups
of nodes where flows are trapped more strongly over time $t$ than one
would expect at random, we find a partition $\mathcal P$ that
maximises
\begin{equation} \label{eq:stability}
    r(t,H_\mathcal{P})=\mbox{trace }R(t, H_\mathcal{P}). 
\end{equation}   
We define $r(t,H_\mathcal{P})$ as the \textit{Markov Stability} of
partition $\mathcal{P}$ at time $t$~\cite{Delvenne2010, Lambiotte2014}.

Maximising $r(t,H_\mathcal{P})$ over the space of all partitions for
each time $t$ results in the sequence of optimal partitions:
\begin{align}
\label{eq:opto}
\mathcal{P}_\text{max}(t) = \operatorname*{arg\,max}_{\mathcal{P}} \,
r(t,H_\mathcal{P}).
\end{align}
Although the optimisation~\eqref{eq:opto} is NP-hard, there exist
efficient heuristic algorithms that work well in practice. In
particular, it has been shown that this optimisation can be carried
out using any algorithm devised for modularity
maximisation~\cite{Delvenne2010, Delvenne2013, Lambiotte2014}.  In
this work, we use the Louvain algorithm~\cite{Blondel2008}, which is
known to offer high quality solutions whilst remaining computationally
efficient. 
The code for Markov Stability 
can be found at \url{github.com/michaelschaub/PartitionStability}.

As an additional improvement of the optimisation of
$\mathcal{P}_\text{max}(t)$, we run the Louvain algorithm $\ell=100$
times with different random initialisations for each Markov time $t$,
and generate an ensemble of solutions
$\{\mathcal{P}_i(t)\}_{i=1}^\ell$.  From this ensemble, we pick the
best partition $\widehat{\mathcal{P}}(t)$ according to our
measure~\eqref{eq:stability}:
$$ \max_i \, \{\mathcal{P}_i(t)\}_{i=1}^\ell  \longmapsto \widehat{\mathcal{P}}(t) \approx \mathcal{P}_\text{max}(t).$$ 
Ideally, the optimised partition from the ensemble,
$\widehat{\mathcal{P}}(t)$, will be close to the true optimum,
$\mathcal{P}_\text{max}(t)$.

To identify the important partitions across time, we use the following
two robustness criteria~\cite{Schaub2012,amor2014}:
\paragraph{\textbf{Consistency of the optimised partition:}}
A relevant partition should be a robust outcome of the optimisation,
i.e., the ensemble of $\ell$ optimised solutions should be similar.
To assess this consistency, we employ an information-theoretical
distance between partitions: the normalised variation of information
between two partitions $\mathcal P$ and $\mathcal P'$ defined
as~\cite{Meila2007}:
\begin{equation}
\label{eq:VI}
 \textit{VI} (\mathcal P, \mathcal P') = \dfrac{2 \Omega(\mathcal P,
 \mathcal P') - \Omega(\mathcal P) - \Omega(\mathcal P')}{\log(n)},
\end{equation}
where $\Omega(\mathcal P) = -\sum_{\mathcal C} p(\mathcal C) \log
p(\mathcal C)$ is a Shannon entropy, with $p(\mathcal C)$ given by the
relative frequency of finding a node in community $\mathcal C$ in
partition $\mathcal P$; $\Omega(\mathcal P, \mathcal P')$ is the
Shannon entropy of the joint probability; and the factor $\log(n)$
ensures that the measure is normalised between $[0,1]$.

To quantify the robustness to the optimisation, we compute the average
variation of information of the ensemble of solutions obtained from
the $\ell$ Louvain runs at Markov time $t$:
\begin{align}
   \text{\VIt} = \dfrac{1}{\ell (\ell-1)} \sum_{i\neq j} \VI (\mathcal{P}_i(t),\mathcal{P}_j(t)).
\end{align}
If all runs of the optimisation return very similar partitions, then
\VIt will be small, indicating robustness of the partition to the
optimisation.  Hence we select partitions with low values (or dips) of
\VIt.

\paragraph{\textbf{Persistence of the partition over time:}}
Relevant partitions should also be optimal across stretches of Markov
time.  Such persistence is indicated both by a plateau in the number of
communities over time and a low value plateau of the cross-time
variation of information:
\begin{align}
    \text{\VItt} = \VI (\widehat{\mathcal{P}}(t),\widehat{\mathcal{P}}(t')).
\end{align}  

Therefore, within a time-block of persistent partitions we 
choose the most robust partition, i.e., that with lowest \VIt.  

\subsection{Quantifying the disruption of community structure under node deletion}
\label{sec:deletion}
To mimic \textit{in silico} the ablation of neuron $i$, we
remove the $i$-th row and column of the adjacency matrix $A$, and
analyse the change induced in the Markov Stability community structure
of the reduced $(n-1)\times (n-1)$ matrix $A_{[i]}$.  Double ablations
are mimicked by simultaneously removing two rows (and their
corresponding columns) to obtain the reduced $(n-2) \times (n-2)$
matrix $A_{[i,j]}$.

\subsubsection{Detecting salient single-node deletions}
\label{sec:single_node}
We carry out a systematic study of all single node deletions in the
network.  To detect relevant deletions, we monitor either an induced
loss of robustness or an induced disruption in the make-up of
particular partitions.

\paragraph{\textbf{Changes induced in the robustness of partitions:}}
First, we run the MS analysis on \textit{all} deletions to obtain the optimised partitions
and their robustness across all times $t$: 
\begin{align}
\left \{\widehat{\mathcal{P}}_{[i]}(t), \, \langle \VI_{[i]}(t)\rangle \right \}_{i=1}^n \quad \forall \, t.
\end{align}
We then fit a Gaussian Process (GP)~\cite{Rasmussen2005} to the
ensemble of $n+1$ time series of the robustness measure $\langle
\VI_{[i]}(t)\rangle$, plus the unablated \VIt.  The resulting GP,
with mean $\mu(t)$ and variance $\sigma^2(t)$, describes the average
robustness of partitions under a single-node deletion.

To detect single-node deletions that induce a large change in the
robustness of a given partition we find sustained outliers of the GP.
For a partition $\widehat{\mathcal{P}}$ optimal over
$t \in [t_1, t_2]$, we select node deletions $i$ such that
$\langle \VI_{[i]}(t)\rangle$ differs from $\mu(t)$ by at least two
standard deviations $\sigma(t)$ over a continuous time interval 
larger than $\ln({\sqrt{t_2/t_1}})$~\cite{amor2014}.  This
criterion identifies node deletions that disrupt the robustness of a
partition over its epoch.

\paragraph{\textbf{Changes induced in the make-up of partitions:}}
To detect if the deletion of node $i$ induces a change in the make-up of partition $\widehat{\mathcal{P}}$, 
we compute the \textit{community variation}:
\begin{equation}
\label{eq:CV}
  \textit{CV}_{[i]} (\widehat{\mathcal{P}})=
  \min_\tau \, \VI (\widehat{\mathcal{P}},\widehat{\mathcal{P}}_{[i]}(\tau)),
\end{equation} 
i.e., the variation of information between $\widehat{\mathcal{P}}$ and
the most similar among \textit{all} optimal partitions of the ablated
network $\widehat{\mathcal{P}}_{[i]}(t)$.

We detect outliers in \CV~for each partition using a simple criterion
based on the inter-percentile range: the deletion of $i$ is
considered an outlier if $\CV_{[i]}>P_{90}+\textit{IPR}_{90/10}$,
where $P_{90}$ is the $90$th percentile, 
and $\textit{IPR}_{90/10}= \abs{P_{90}-P_{10}}$ is
the interpercentile range between the 10th-90th percentiles of
the ensemble of $\CV_{[i]}$.

\subsubsection{Detecting supra-additive double-node deletions}  
\label{sec:qr}
We have carried out a study of all double deletions in the network to detect 
two-node deletions whose effect is larger than the additive effect of the 
two corresponding single node deletions. 
To this end, we first obtain the set of MS partitions across all Markov times 
for all double delections $\widehat{\mathcal{P}}_{[i,j]}(t)$,
and compute their community variation:
\begin{equation}
  \textit{CV}_{[i,j]}(\widehat{\mathcal{P}})=
  \min_\tau \, \VI (\widehat{\mathcal{P}},\widehat{\mathcal{P}}_{[i,j]}(\tau)).
\end{equation}
We then compute the average of the individual ablations: 
\begin{align}
\label{eq:CV_i_j}
\textit{CV}_{[i],[j]}(\widehat{\mathcal{P}}) = \dfrac{CV_{[i]}(\widehat{\mathcal{P}}) + CV_{[j]}(\widehat{\mathcal{P}})}{2}.
\end{align}

To find pairs with a supra-additive effect, we use Quantile Regression (QR)~\cite{Koenker2005}, 
a method widely used in econometrics, ecology, and medical statistics.
Whereas least squares regression aims to estimate the conditional mean of the samples,
QR provides a method to estimate conditional quantiles of the sample distribution.
Hence, QR facilitates a more global representation of the relationships between the 
dependent and independent variables considered in the regression.
A good introduction to QR can be found in Ref.~\cite{Cade2003}, and
a more in-depth treatment can be found in the book by Koenker~\cite{Koenker2005}.

For a partition $\widehat{\mathcal{P}}$, we employ QR to fit quantiles for the regression of
$\textit{CV}_{[i,j]}(\widehat{\mathcal{P}})$ against $\textit{CV}_{[i],[j]}(\widehat{\mathcal{P}})$,
using all 38781 two-node ablations (Fig.~\ref{fig:QR}).  
We report the top 1\% double deletions according to their quantile scores---this is our criterion to select double-ablations that have a strong effect. 
All scores are computed using Bayesian Quantile Regression, as implemented in the 
R~package~\texttt{BSquare} (\url{https://cran.r-project.org/web/packages/BSquare/index.html}), 
which fits all quantiles simultaneously resulting in a more coherent estimate~\cite{Smith2013}.   
Following Ref.~\cite{Smith2013},  we fit the quantiles to the normalised $\textit{CV}_{[i],[j]}(\widehat{\mathcal{P}})$
using a Gamma centering distribution and four basis functions.

\subsection{Finding flow roles in networks: Role-Based Similarity}
\label{sec:RBS}

In directed networks, nodes can have different `roles', e.g., sinks,
sources or hubs.  In complex directed networks, functional roles may
not fall into such simple categories, yet nodes can still be
characterised by their contribution to the diffusion of in- and
out-flows.  Here we use a recent method (Role-Based Similarity, RBS)
to uncover roles in directed networks based on the patterns of
incoming and outgoing flows at all scales~\cite{Cooper2010,
  Cooper2010a}.  The main idea underpinning RBS is that nodes with a
similar in/out flow profile play a similar role, regardless of whether
they are near or far apart in the network.  Each node is associated
with a feature vector $ \vec{x}_i$ containing a weighted number of in-
and out-paths of increasing lengths beginning and ending at the
node. The feature vectors are collected in the feature matrix $X$:
\begin{equation} \label{flowprofile} 
    X= \left[ 
        \begin{matrix} \vec{x}_1\\ \vdots \\ \vec{x}_n 
        \end{matrix} \right ] = \overbrace{ \left[ \dots (\beta
            A^{T})^k \vec{1} \dots \right. }^{\text{paths in}}| 
    \overbrace{\left. \dots (\beta A )^k 
            \vec{1} \dots \right] }^{\text{paths out}}, 
\end{equation} 
where $\beta=\alpha/\lambda_1$, with $\lambda_1$ the spectral radius of the adjacency matrix $A$
and $\alpha\in(0,1)$.  The cosine between feature vectors gives the similarity score between nodes:
\begin{equation} 
    Y_{ij}=
    \frac{\vec{x}_i\vec{x}_j^{T}}{{\norm{\vec{x}_i}}_2 \, {\norm{\vec{x}_j}}_2}.  
\end{equation}
The $n\times n$ matrix $Y$ quantifies how similar the
directed flow profiles between every pair of nodes are.  
Nodes with identical connectivity have $Y_{ij}=1$, whereas in the 
case of nodes with dissimilar flow profiles
(e.g., if $i$ is a source node with no incoming connections and $j$ is
a sink node with no outgoing connections), then their feature
vectors are orthogonal and $Y_{ij}=0$.

As outlined in Refs.~\cite{Cooper2010,Cooper2010a,Beguerisse2013}, we
compute the similarity matrix $Y$ iteratively with $\alpha=0.95$, and
apply the RMST algorithm to obtain a \textit{similarity graph}, in
which only the important information of $Y$ is retained. We then
extract \textit{flow roles} in a data-driven manner without imposing
the number of roles \textit{a priori} by clustering the similarity
graph (see Fig.~\ref{S4_Fig}).  The flow roles so obtained have been
shown to capture relevant features in complex networks, where other
role classifications based on combinatorial concepts and
neighbourhoods fail~\cite{Beguerisse2013,Beguerisse2014}. In
particular, our flow roles are fundamentally different from notions of
roles in social networks based on Structural
Equivalence~\cite{Lorrain1971} and Regular
Equivalence~\cite{Everett1994}. Such equivalence measures do not
incorporate information about the large scales of the network and are
sensitive to small perturbations, making them unsuitable for complex
networks such as the \CE~connectome~\cite{Beguerisse2014} (see Fig.~\ref{S6_Fig} for roles based on Regular Equivalence).

\subsection{Acknowledgements}
KAB acknowledges an Award from the Imperial~College Undergraduate
Research Opportunities Programme (UROP).  MTS acknowledges support
from the ARC and the Belgium network DYSCO (Dynamical Systems, Control
and Optimisation).  YNB acknowledges support from the G.~Harold and
Leila Y.~Mathers Foundation.  MBD acknowledges support from the James
S. McDonnell Foundation Postdoctoral Program in
Complexity~Science/Complex~Systems Fellowship Award (\#220020349-CS/PD
Fellow).  MB acknowledges support from EPSRC grant EP/I017267/1 under
the Mathematics Underpinning the Digital Economy program.

\appendix
\onecolumngrid

\counterwithin{figure}{section}
\setcounter{section}{0}
\setcounter{equation}{0}
\renewcommand\thesection{}
\renewcommand\thesubsection{SI\arabic{subsection}}
\renewcommand\thefigure{S\arabic{figure}}
\renewcommand\theequation{S\arabic{equation}}

\section{Supplementary Information}

\subsection{Supplementary Data}
\label{S1_Data}
{\bf Supplementary Data as XLS spreadsheet}  is available at \url{     http://dx.doi.org/10.1371/journal.pcbi.1005055
}.

\subsection{Supplementary Text 1}
\label{S1_Text}
\textbf{Comparison of MS partitions to other methods}

The flow-based MS partitions are distinct from
partitions obtained by several other methods.  In particular, we have
compared against partitions obtained with Modularity, Stochastic Block
models, and Infomap.

Modularity has been used to obtain optimised partitions in
Refs.~\cite{pan2010,sohn2011}.  The partition found in
Ref.~\cite{pan2010} is closest to our 4-way Partition $\mathcal{B}$
(\textit{VI} = 0.185), whereas the partition found in
Ref.~\cite{sohn2011} is closest to our 3-way Partition $\mathcal{C}$
(\textit{VI} = 0.186).  Note that optimisation of modularity at a
fixed resolution imposes an intrinsic scale, so that partitions found
with modularity are well matched to a particular scale (i.e., a
particular Markov time) in the Markov Stability framework, as shown
previously~\cite{Delvenne2010,Lambiotte2014}.  On the other hand, as
discussed in the main text, the Markov Stability framework carries out a
systematic scanning across Markov times~\cite{Schaub2012} allowing the
intrinsic multiscale organisation to became apparent.

The partitions based on stochastic block models~\cite{pavlovic2014}
and hierarchical Infomap~\cite{Edler2015} are less similar to the ones
found by MS: the partition found by stochastic block models
in~\cite{pavlovic2014} is closest to our 3-way Partition $\mathcal{C}$
(but with a higher \textit{VI}=0.272), and the partition found by
hierarchical Infomap in~\cite{Edler2015} is closest to our 6-way
Partition $\mathcal{A}$ (yet with an even higher
\textit{VI}=0.282). These differences in the outcomes are expected due
to the contrasting methodological approaches.  In particular, Infomap
is known to impose a clique-like structure to the modules leading to
groupings where strong local density is favoured~\cite{Schaub2012a}.
We remark that, as shown in S2 Text, the MS communities
are also different from the flow roles found through RBS.

\subsection{Supplementary Text 2}
\label{S2_Text}
\textbf{Comparison of RBS flow roles with other analyses of roles}

Our RBS \textit{flow roles} are fundamentally different from notions of
roles used in social networks based on Structural Equivalence
(SE)~\cite{Lorrain1971}, and Regular Equivalence
(RE)~\cite{Everett1994} (Fig.~S6 and S1 Data).
Because both RE and SE consider only one-step neighbourhoods and do
not incorporate information about the long scales of the network
\cite{Schaub2012}, they are less applicable to complex networks such
as the \CE~connectome~\cite{Beguerisse2014}.  In particular the roles
produced by REGE show undifferentiated PageRank and connectivity
profiles.

In Refs.~\cite{pan2010,sohn2011}, roles were assigned to neurons
according to `community roles', the technique proposed by Guimera et
al~\cite{Guimera2005} which identifies certain interneurons as relevant
hubs between predefined communities.  It was found that command
interneurons (e.g. AVA, AVB, AVD, PVC) play the role of global hubs,
whereas D-type motor neurons play the role of provincial hubs~\cite{pan2010}.
These features are in line with our ablation results, where D-type motor
neuron ablations alter flows at finer scales and ablation of
interneurons modifies flow patterns at larger scales. 
Indeed, the concept of 'community role' is closer to our ablation results, in that we
measure there the disruption of flow-based communities. 

Chatterjee and Sinha~\cite{chatterjee2007} explored the core-periphery
structure of the \textit{C. elegans} connectome using a $k$-core
decomposition based on in- and out- degree separately.  The $k$-core
of a network is the subgraph with the property that all nodes have
(in/out)degree at least $k$.  As expected, motor neurons are
overrepresented in the $k$-cores based on in-degree, and sensory
neurons are overrepresented in $k$-cores based on out-degree.  This
distinction between neurons with upstream and downstream roles is also
an inherent characteristic in the RBS analysis, yet from a different
perspective, i.e., based on the global characteristics of a node with
respect to the in- and out-flows in the network, rather than based on
its local connections.

To quantify the differences between the groupings into roles obtained by these
different methods, we have computed the variation of information between them.
RBS roles show very low similarity to any of the other groupings with values 
of $\textit{VI}=0.3061, 0.3607, 0.4356$ against REGE~\cite{Borgatti1993},
Sohn et al.~\cite{sohn2011} and Pan et al.~\cite{pan2010}, respectively.

\clearpage

\begin{figure}[tp]
    \centering
    \includegraphics[width=.9\textwidth]{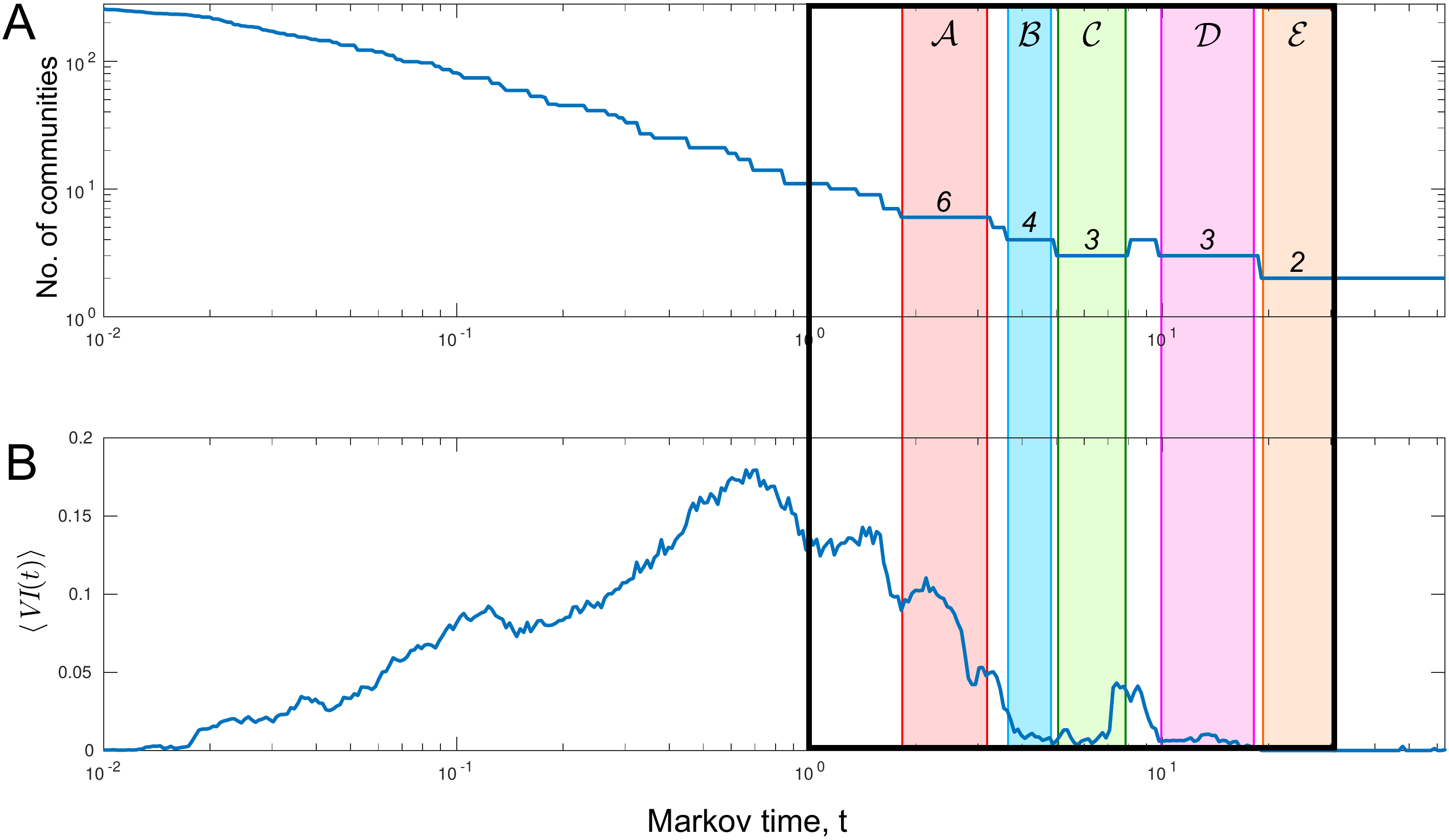} 
    \caption{\textbf{Full analysis of the \CE~connectome with Markov Stability (MS)}. 
We show the scan across all Markov times, from the finest possible partition (every node in its own partition) at small Markov times to the bipartition at large Markov times. The highlighted time interval corresponds to Fig. 1 in the main text, which focusses on the medium to coarse partitions $\mathcal{A}-\mathcal{E}$.}
\label{S1_Fig}
\end{figure}

\begin{figure}[tp]
    \centering
    \includegraphics[width=.9\textwidth]{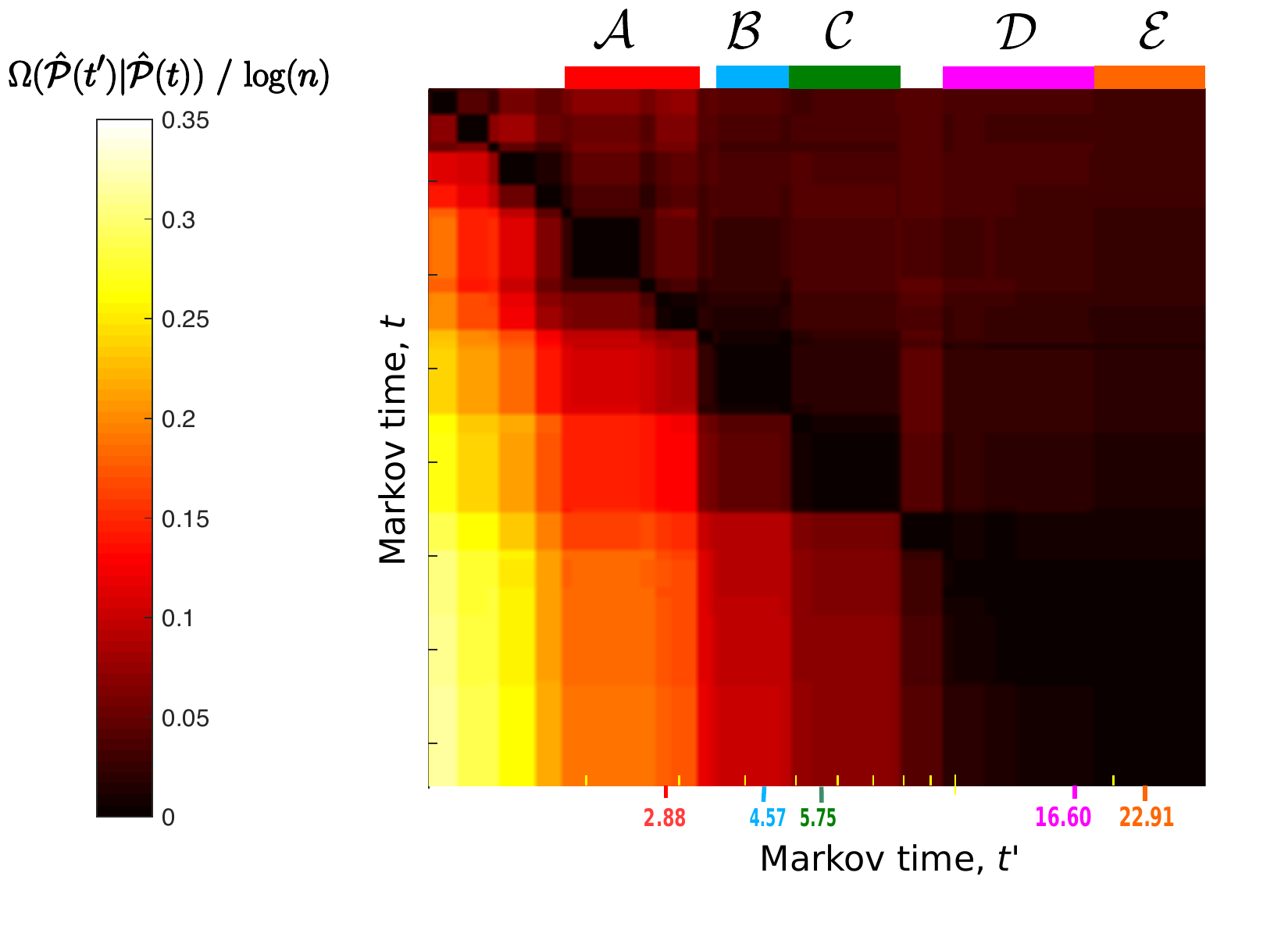} 
    \caption{\textbf{The asymmetry in the normalised conditional entropy of the optimised MS
partitions signals a quasi-hierarchical community structure.}
The normalized conditional entropy $\Omega(\mathcal P(t')|\mathcal P(t))/ \log(n) \in [0,1]$ 
quantifies the uncertainty in the community assignment
$\mathcal P(t')$ given the known partition $\mathcal P(t)$.  If
$\mathcal P(t')$ can be predicted from $\mathcal P(t)$,
(i.e. when $\mathcal P(t')$ is a strictly hierarchical agglomeration of
the communities of $\mathcal P(t)$) then the conditional entropy will
be zero.  The strong upper-triangular character of the conditional entropy
of the partitions $\mathcal{A}-\mathcal{E}$ indicates a quasi-hierarchical organisation. 
}
\label{S2_Fig}
\end{figure}

\begin{figure}[tp]
    \centering
    \includegraphics[width=.6\textwidth]{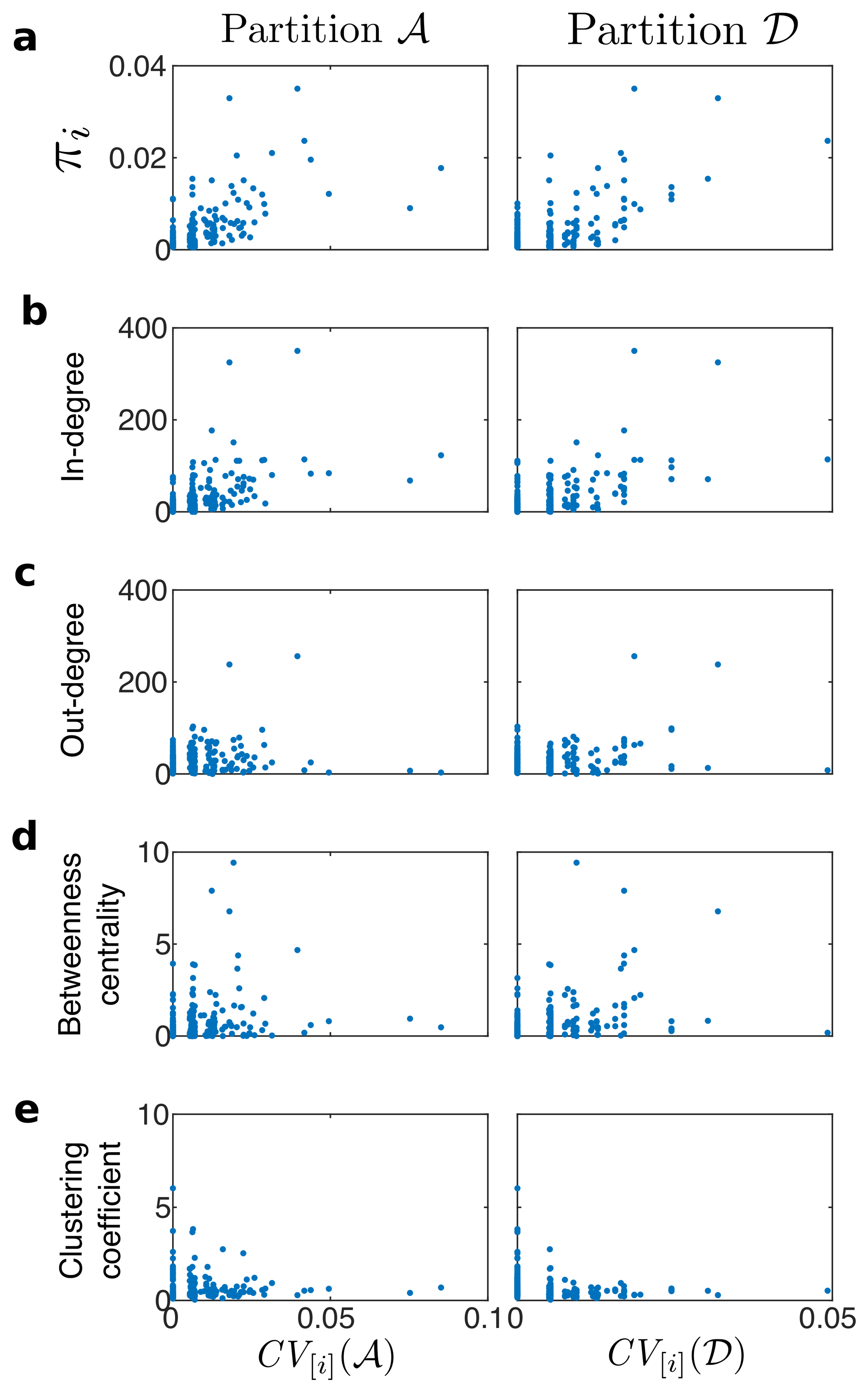} 

\caption{\textbf{The effect of ablations and other network measures.}
Scatter plots of the Community Variation with respect to Partitions $\mathcal{A}$ and $\mathcal{D}$, $\textit{CV}_{[i]}(\mathcal{A})$ (left column) and $\textit{CV}_{[i]}(\mathcal{D})$ (right column), for all single neuron ablations ($i=1,\ldots, 279$) plotted against the following properties of the corresponding neuron: {\bf a}, stationary flow distribution $\pi$ (PageRank); {\bf b}, in-degree; {\bf c}, out-degree; {\bf d}, betweenness centrality;  and {\bf e}, local clustering coefficient. None of these quantities (which are related to network centralities) shows a manifest correlation with the effect of the neuron ablation on community structure. } 
\label{S3_Fig}
\label{fig:sal_ablations}
\end{figure}

\begin{figure}[tp]
    \centering
    \includegraphics[width=.9\textwidth]{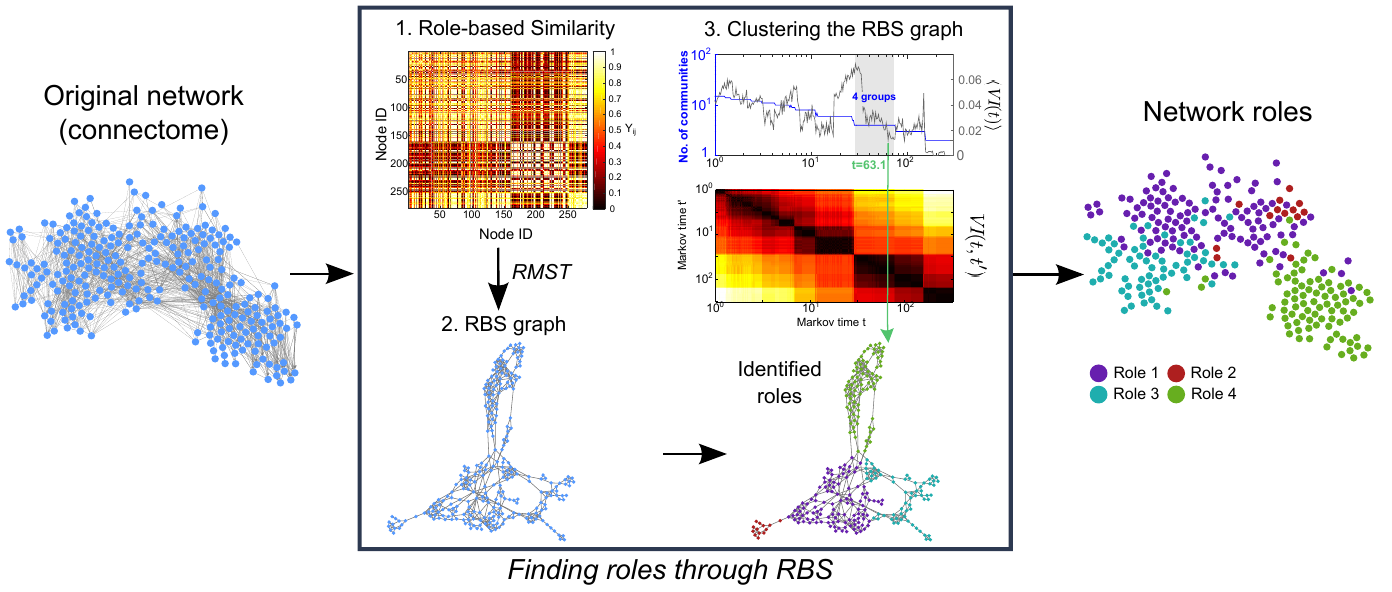} 
   \caption{ \textbf{Finding role profiles with RBS.}
    Schematic summary of the procedure to obtain flow roles using RBS analysis, as discussed in detail in \cite{Beguerisse2013}. First, from the original {\it directed} network of the {\it C. elegans} connectome we create a similarity matrix using the RBS metric, by computing a similarity score between each node in the network, based on their incoming and outgoing weighted path profiles. Second, the similarity matrix is transformed into a similarity matrix using the RMST method, which subsequently prunes out uninformative links (see Ref. \cite{Beguerisse2013} for details).  Third, the resulting similarity graph is clustered to obtain relevant groups of nodes with similar in- and out-flow profiles at all scales. Four such classes of neurons (\textit{flow roles}) are found in this case.  The neurons are then colored according to their flow profile on the original connectome layout. }
\label{fig:rbs_si}
\label{S4_Fig}   
    \end{figure}

\begin{figure}[tp]
    \centering
    \includegraphics[width=.9\textwidth]{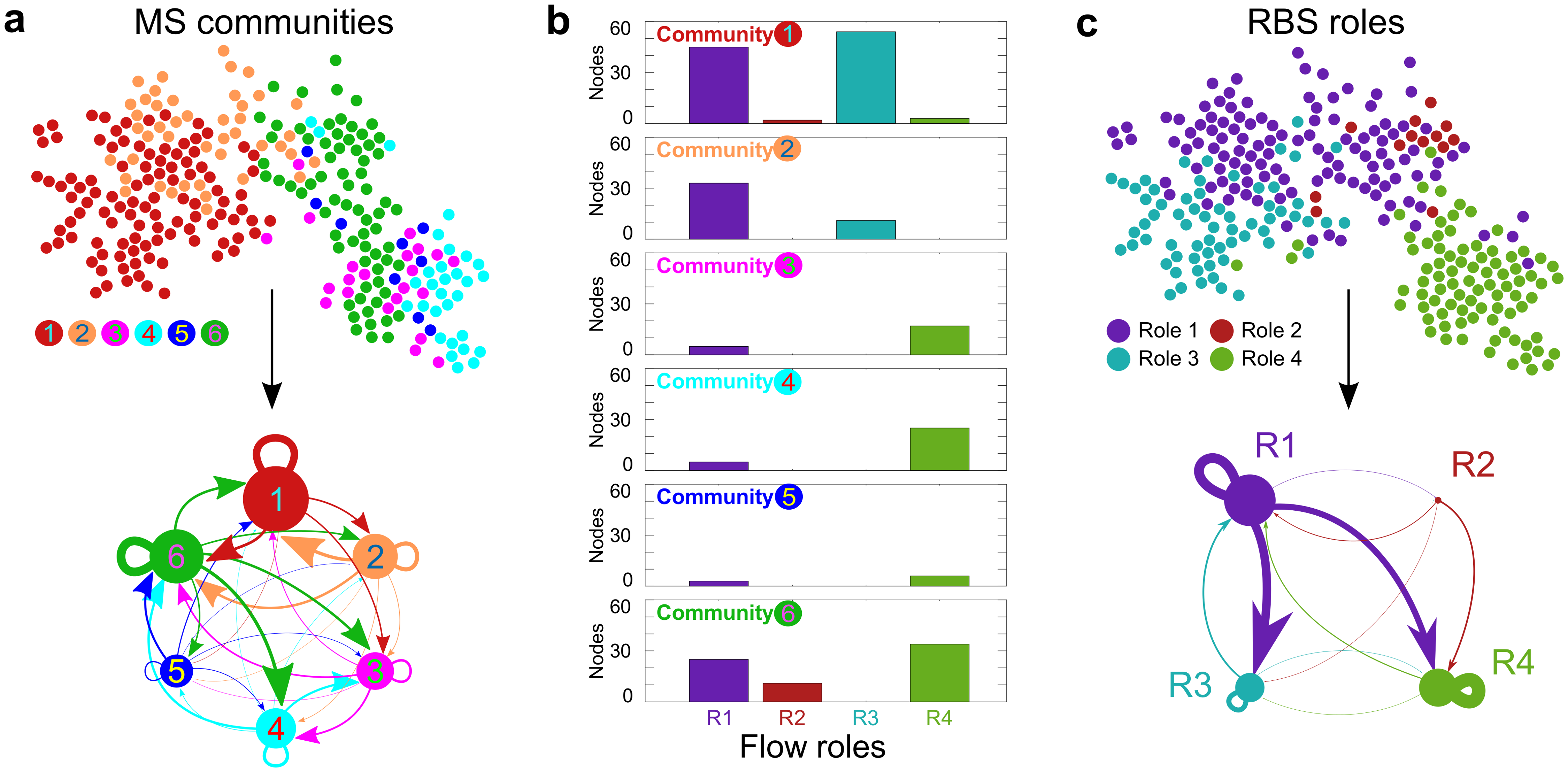} 

\caption{\textbf{Distribution of RBS flow roles across MS communities.}
RBS roles in each of the six communities of partition $\mathcal{A}$.  The communities and flow roles induce very different groupings in the connectome.  Hence the six communities present distinct mixes of roles: the anterior communities $\mathcal{A}1$ and $\mathcal{A}2$ present a dominance of roles R1 and R3, whereas the posterior communities $\mathcal{A}3$, $\mathcal{A}4$ and $\mathcal{A}5$ are dominated by roles R1 and R4. Community $\mathcal{A}6$ has a balanced mix of roles R1, R2, and R4 giving it a distinctive information processing structure, confirming the the importance of its embedded rich-club neurons. }
\label{fig:organigram}
\label{S5_Fig}
\end{figure}

\begin{figure}
    \centering
    \includegraphics[width=.9\textwidth]{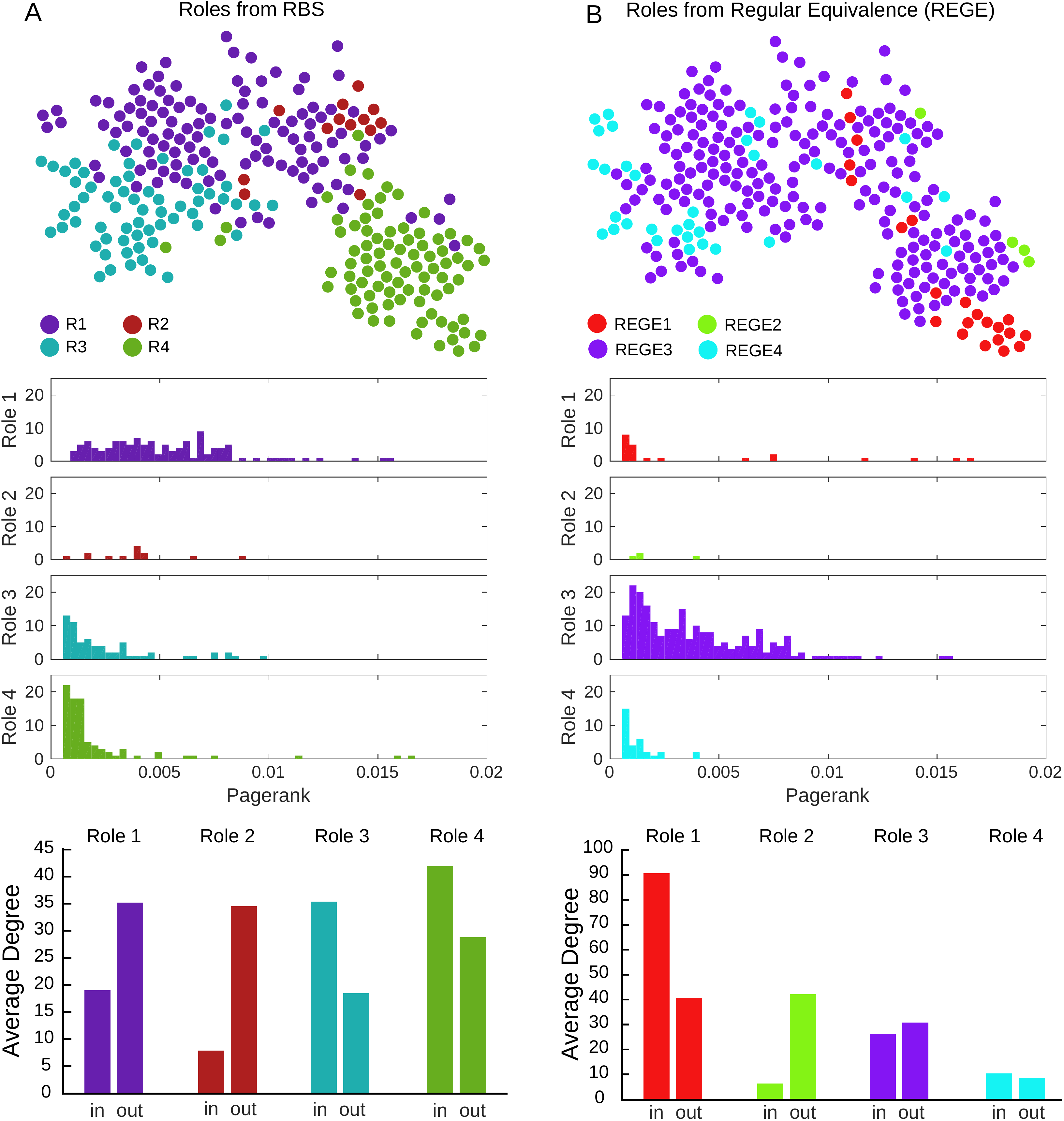} 

\caption{\textbf{Comparison of RBS flow roles to roles obtained using Regular Equivalence.}
{\bf a}: Roles of the nodes according to RBS with the PageRank distribution for each role and the average in/out degree for each role. {\bf b}: Same for the roles obtained according to Regular Equivalence obtained using the REGE algorithm~\cite{Borgatti1993}.}
\label{fig:organigram2}
\label{S6_Fig}
\label{fig:REGE}
\end{figure}

\begin{figure}
\centering
\includegraphics[height=0.9\textheight]{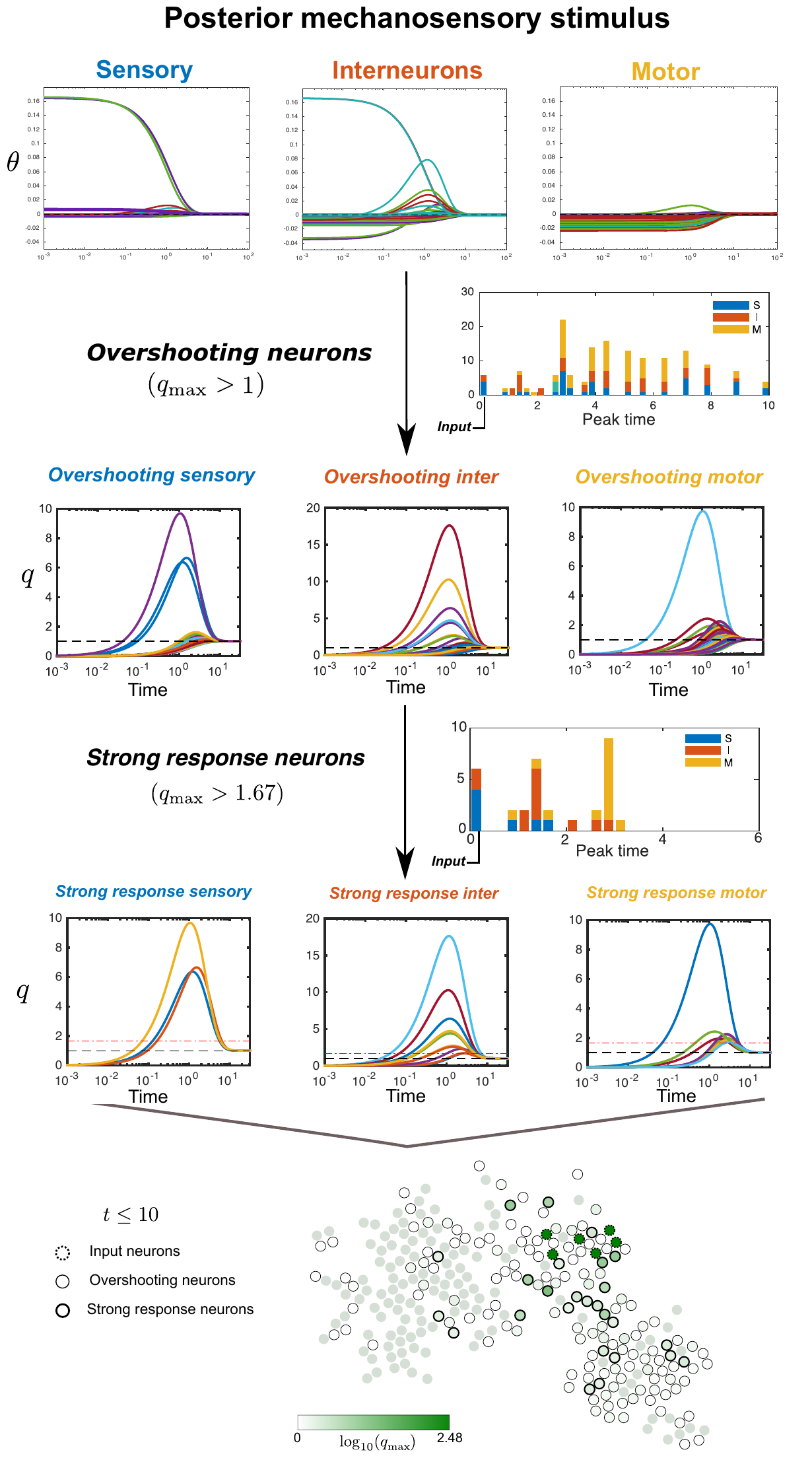} 

\caption{\textbf{Summary of the procedure for signal propagation analysis of posterior mechanosensory stimulus scenario (i1).}
For all neurons, we compute $\phi_i(t)$, i.e., the amount of signal present at each node at Markov time $t$. 
As time grows, the signal at each node converges to its stationary value $\pi_i$. 
Hence $\theta_i(t) = \phi_i(t) - \pi_i \rightarrow 0$.  The approach to stationarity can happen in two ways: i) the initially negative $\theta_i(t)$ approaches 0 from below; ii) $\theta_i(t)$ 'overshoots' before decaying towards its stationary value.  We consider the signal relative to the stationary value, $q_i(t) = \phi_i(t)/\pi_i$, and focus on neurons that overshoot (i.e., those with $q_{\text{max},i} := \max_t \phi_i(t)/\pi_i > 1$) and we collect the times at which they reach their peak.  A concise summary of the signal propagation is given by the \textit{strong response neurons} with $q_\text{max,i} > 5/3$.  Their peak-time histogram and the particular sequence of strong
response neurons is characteristic of the different input-response biological scenarios, as well as the analyses by neuron type and flow roles.  }
\label{fig:How}
\label{S7_Fig}
\end{figure}

\begin{figure}
\centering
\includegraphics[width=.9\textwidth]{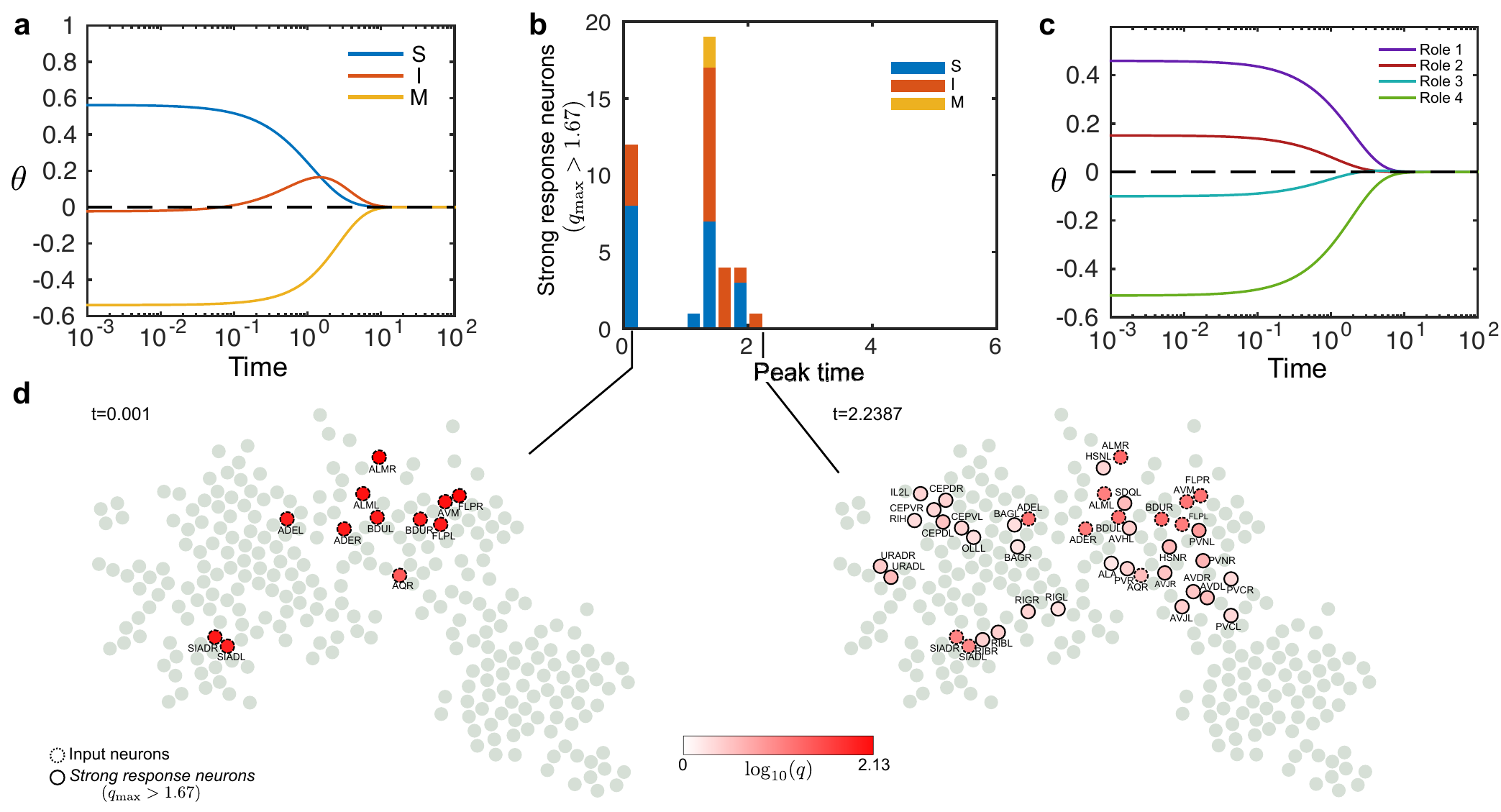} 

\caption{    \textbf{Signal propagation of the anterior mechanosensory stimulus (i2).} 
    Signal propagation evolving from an initial condition localised at the mechanosensory neurons (i2).  \textbf{(a)} As stationarity is approached ($\boldsymbol{\theta}(t) \to 0$), the input propagates from sensory to motor neurons through an intermediate stage when interneurons overshoot.  \textbf{(b)} The propagation seen as a cascade of strong response neurons ( $q_\text{max,i}> 1+2/3$) with peak times concentrated around two bursts.  \textbf{(c)} The input (i2), appears localised on R1 and to a lesser extent R2 neurons. The signal diffuses somewhat quicker out of R2 than R1 neurons, but induces not collective overshoot of R3 or R4 neurons.  \textbf{(d)} Stages of signal propagation in the network showing the strong response neurons that have peaked at each time.  }
    \label{S8_Fig}
    \label{fig:Summary1}
\end{figure}

\begin{figure}
\centering
\includegraphics[width=.9\textwidth]{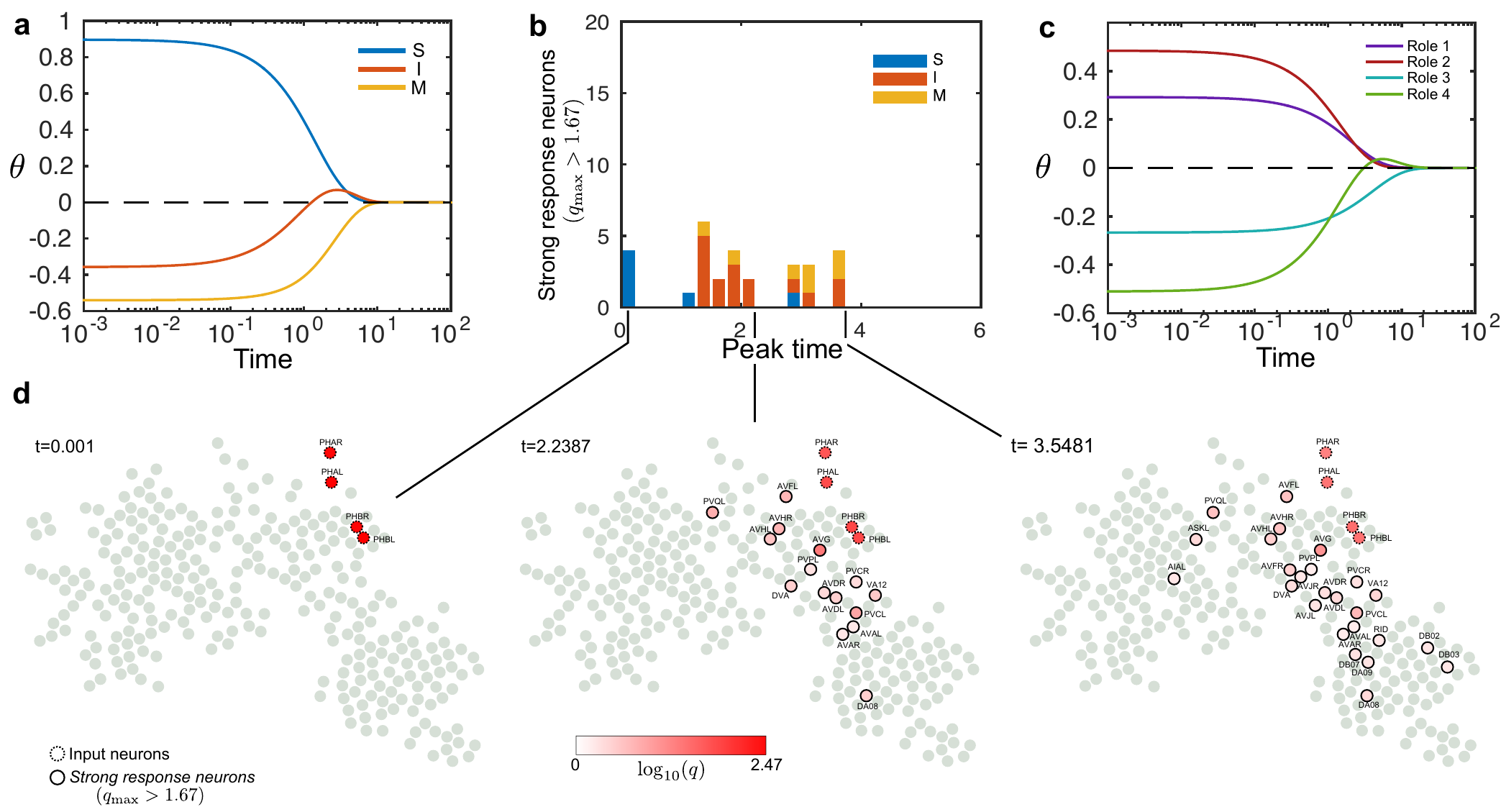} 

 \caption{   \textbf{Signal propagation: posterior chemosensory stimulus (i3).}
    See caption of \ref{fig:Summary1}.  }
    \label{S9_Fig}
    \label{fig:Summary2}
\end{figure}

\begin{figure}
\centering
\includegraphics[width=.9\textwidth]{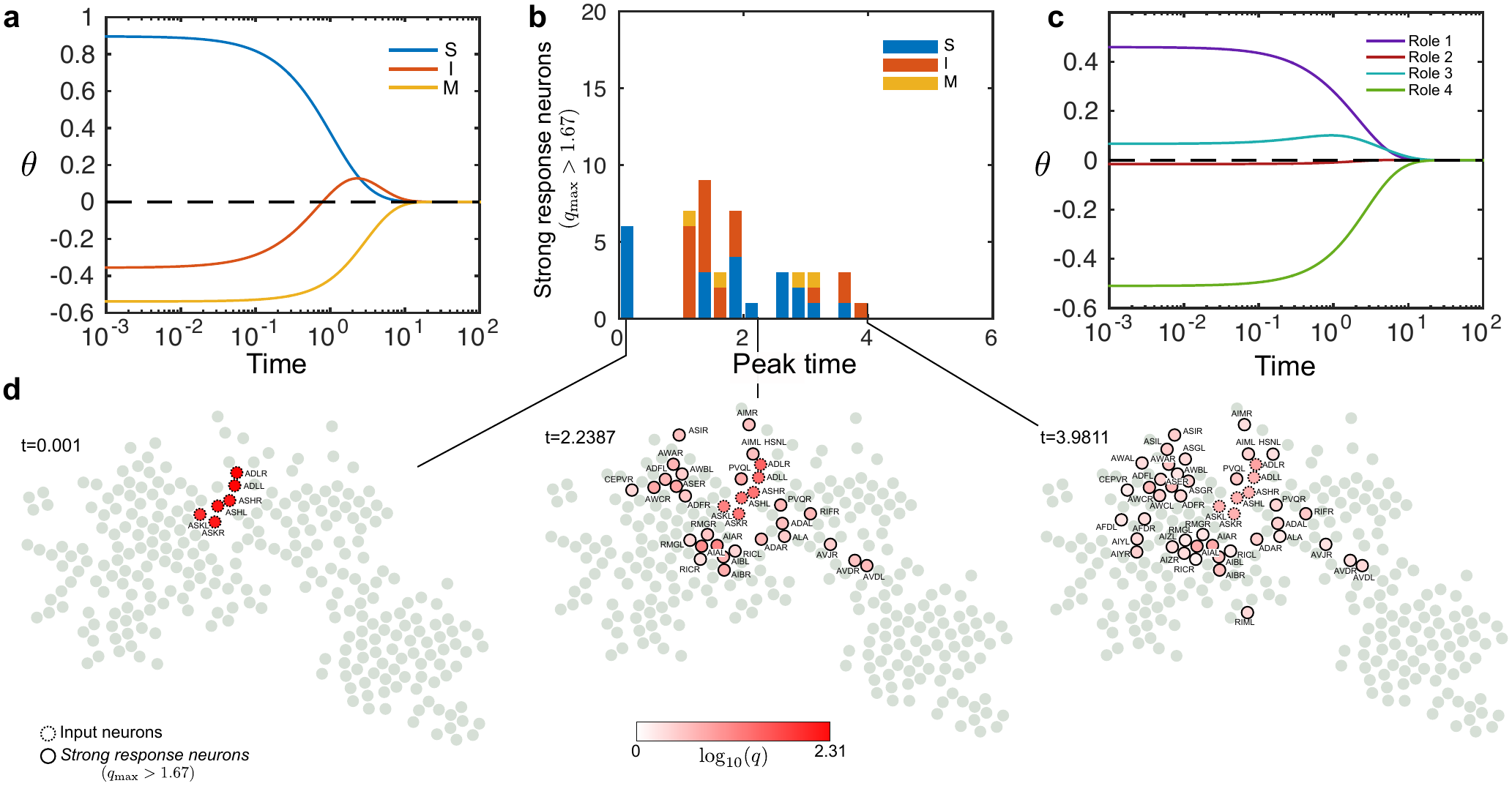} 

    \caption{\textbf{Signal propagation: anterior chemosensory stimulus (i4).}
    See caption of \ref{fig:Summary1}.} 
    \label{S10_Fig}
    \label{fig:Summary3}
\end{figure}

\begin{figure}
\centering
\includegraphics[width=.9\textwidth]{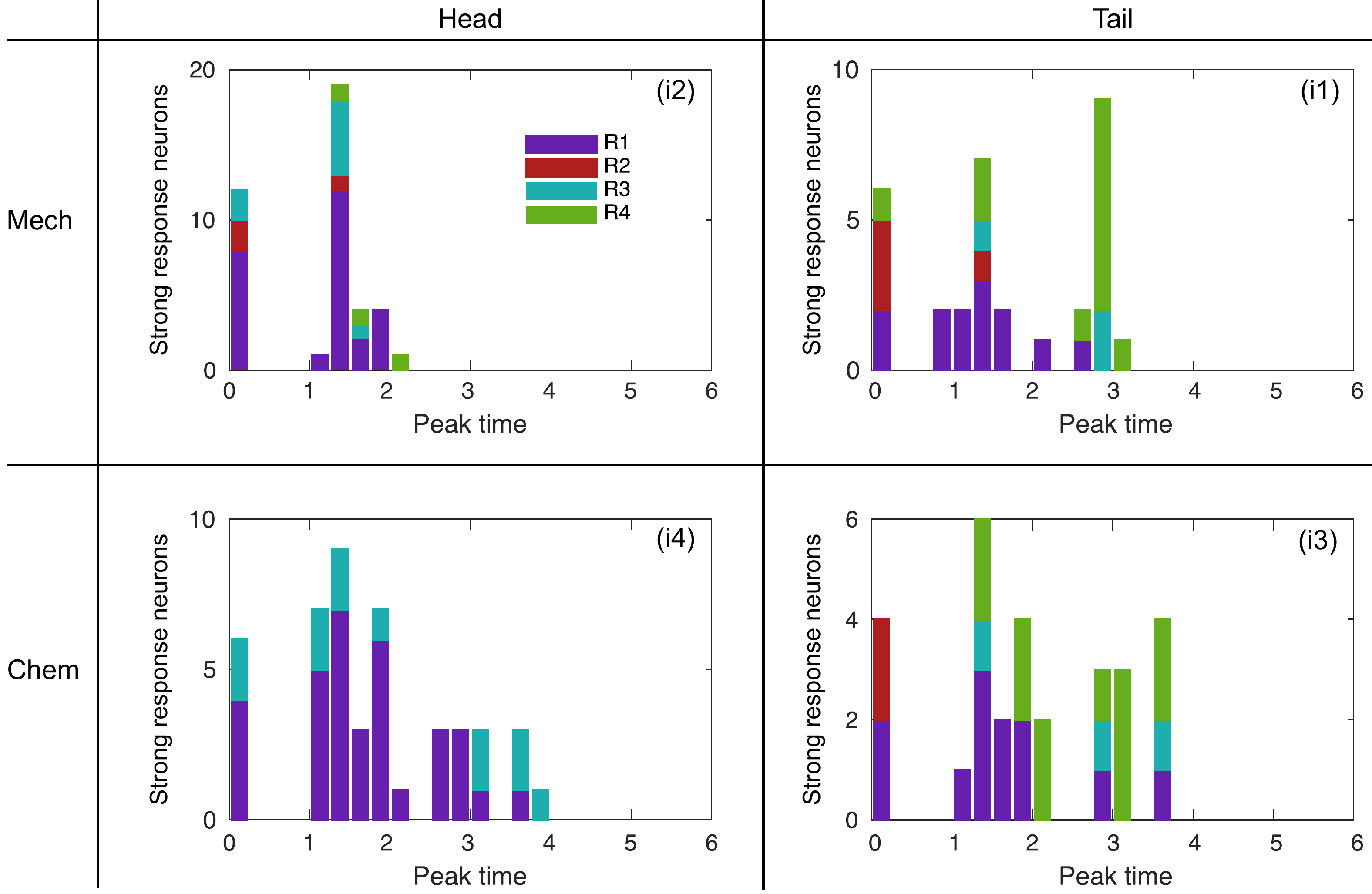} 
\caption{    \textbf{Peak times of strong response neurons by RBS roles for each of the four input scenarios (i1)-(i4).}
    Histograms of peak times of the strong response neurons in the four biological scenarios from the perspective of flow roles. The tail inputs (i1) and (i3) induce strong responses on neurons spreading from R2 to R1 and finally to R4.  On the other hand, the head inputs induce strong responses on neurons heavily based on R1 spreading downwards to R3.  }
    \label{S11_Fig}
    \label{fig:Summary4}
\end{figure}

\begin{figure}
\centering
\includegraphics[width=.35\textwidth]{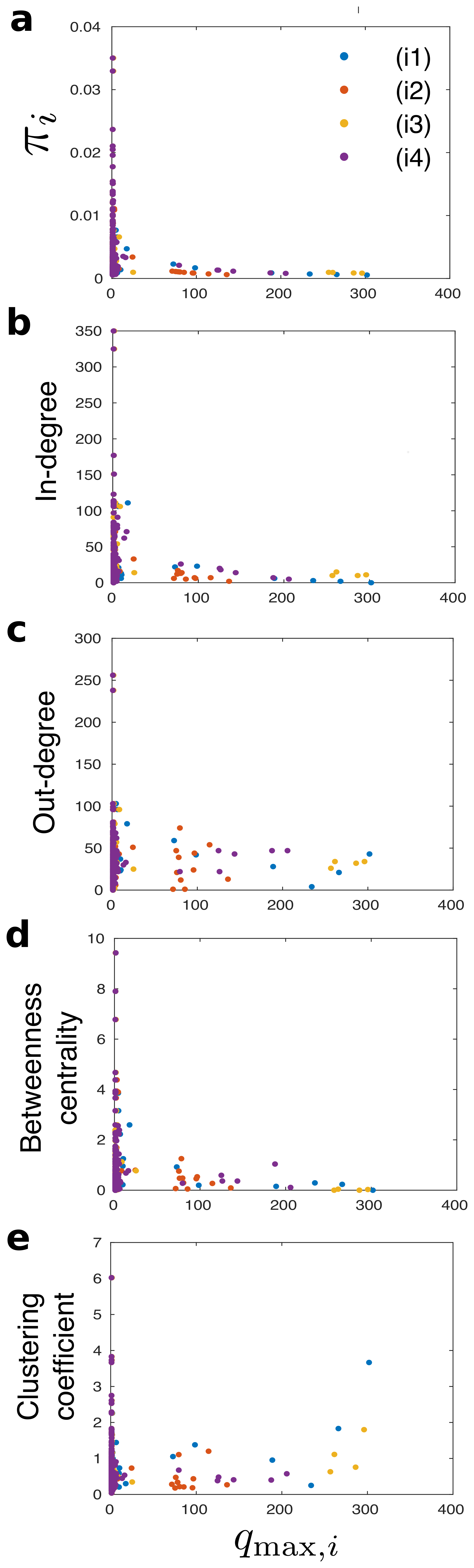} 
\caption{\textbf{Peak overshoots against other network measures.} The maximum overshoot of each neuron $q_{\text{max},i}$ for each of the four biological scenarios (i1)--(i4) is plotted against the following measures of the corresponding neuron: {\bf a}, stationary flow distribution $\pi$ (PageRank); {\bf b}, in-degree; {\bf c}, out-degree; {\bf d}, betweenness centrality;  and {\bf e}, local clustering coefficient. There is no manifest correlation between the overshooting $q_{\text{max},i}$ and any of those centrality scores or the local clustering coefficient.}
\label{S12_Fig}
\label{fig:CentralitySignal}
\end{figure}

\end{document}